\documentclass[useAMS,usegraphicx]{mn2e}
\usepackage{longtable}
\usepackage{aas_macros}
\usepackage{rotating}

\renewcommand{\sc}{\scshape}
\def\herschel{{\it Herschel}}

\def\ergs{\,erg\,cm$^{-2}$\,s$^{-1}$}
\def\gs{\mathrel{\raise0.35ex\hbox{$\scriptstyle >$}\kern-0.6em
\lower0.40ex\hbox{{$\scriptstyle \sim$}}}}
\def\ls{\mathrel{\raise0.35ex\hbox{$\scriptstyle <$}\kern-0.6em
\lower0.40ex\hbox{{$\scriptstyle \sim$}}}}

\title[Near-IR spectroscopy of {\it Herschel} galaxies]
{FMOS near-IR spectroscopy of {\it Herschel} selected galaxies: star formation rates, metallicity and dust attenuation at \textit{z}$\sim1$ }
\author[I.G.~Roseboom et al.]
{\parbox{\textwidth}{\raggedright I.G.~Roseboom,$^{1,2}$\thanks{E-mail: \texttt{igr@roe.ac.uk}}
A.~Bunker,$^{3}$
M.~Sumiyoshi,$^{4}$
L.~Wang,$^{2}$
G.~Dalton,$^{3,5}$
M.~Akiyama,$^{6}$
J.~Bock,$^{7,8}$
D.~Bonfield$^{9}$
V.~Buat,$^{10}$
C.~Casey,$^{11}$
E.~Chapin,$^{12}$
D.L.~Clements,$^{13}$
A.~Conley,$^{14}$
E.~Curtis-Lake,$^{1}$
A.~Cooray,$^{15,7}$
J.S.~Dunlop,$^{1}$
D.~Farrah,$^{2}$
S.J.~Ham,$^{3}$
E.~Ibar,$^{16}$
F.~Iwamuro,$^{4}$
M.~Kimura,$^{17}$ 
I.~Lewis,$^{3}$
E.~Macaulay,$^{3}$
G.~Magdis,$^{3}$
T.Maihara,$^{4}$
G.~Marsden,$^{12}$
T.~Mauch,$^{3,9}$
Y.~Moritani,$^{4}$
K.~Ohta,$^{4}$
S.J.~Oliver,$^{2}$
M.J.~Page,$^{18}$
B.~Schulz,$^{19,8}$
Douglas~Scott,$^{12}$
M.~Symeonidis,$^{18}$
N.~Takato,$^{17}$
N.~Tamura,$^{17}$
T.~Totani,$^{4}$
K.~Yabe,$^{4}$
M.~Zemcov,$^{7,8}$}\vspace{0.4cm}\\
\parbox{\textwidth}{\raggedright $^{1}$Institute for Astronomy, University of Edinburgh, Royal Observatory, Blackford Hill, Edinburgh EH9 3HJ, UK\\
$^{2}$Astronomy Centre, Dept. of Physics \& Astronomy, University of Sussex, Brighton BN1 9QH, UK\\
$^{3}$Astrophysics, University of Oxford, Keble Road, Oxford OX1 3RH, UK\\
$^{4}$Department of Astronomy, Faculty of Science, Kyoto University, Kyoto 606-8502, Japan\\
$^{5}$RALSpace, Rutherford Appleton Laboratory, Chilton, Didcot, Oxfordshire OX11 0QX, UK\\
$^{6}$Astronomical Institute, Tohoku University, Aoba-ku, Sendai, 980-8578, Japan\\
$^{7}$California Institute of Technology, 1200 E. California Blvd., Pasadena, CA 91125, USA\\
$^{8}$Jet Propulsion Laboratory, 4800 Oak Grove Drive, Pasadena, CA 91109, USA\\
$^{9}$Centre for Astrophysics Research, University of Hertfordshire, College Lane, Hatfield, Hertfordshire AL10 9AB, UK\\
$^{10}$Laboratoire d'Astrophysique de Marseille, OAMP, Universit\'e Aix-marseille, CNRS, 38 rue Fr\'ed\'eric Joliot-Curie, 13388 Marseille cedex 13, France\\
$^{11}$Institute for Astronomy, University of Hawaii, Manoa, HI 96822, USA ; Canada-France-Hawaii Telescope Corp., Kamuela, HI 96743, USA\\
$^{12}$Department of Physics \& Astronomy, University of British Columbia, 6224 Agricultural Road, Vancouver, BC V6T~1Z1, Canada\\
$^{13}$Astrophysics Group, Imperial College London, Blackett Laboratory, Prince Consort Road, London SW7 2AZ, UK\\
$^{14}$Center for Astrophysics and Space Astronomy, 593 UCB, Boulder, Co 80309-0593, USA\\
$^{15}$Dept. of Physics \& Astronomy, University of California, Irvine, CA 92697, USA\\
$^{16}$UK Astronomy Technology Centre, Royal Observatory, Blackford Hill, Edinburgh EH9 3HJ, UK\\
$^{17}$Subaru Telescope, NAOJ, 650 North Aohoku Place, Hilo, HI, 96720,USA\\
$^{18}$Mullard Space Science Laboratory, University College London, Holmbury St. Mary, Dorking, Surrey RH5 6NT, UK\\
$^{19}$Infrared Processing and Analysis Center, MS 100-22, California Institute of Technology, JPL, Pasadena, CA 91125, USA\\
}
}
\begin{document}

\date{\today}

\pagerange{\pageref{firstpage}--\pageref{lastpage}} \pubyear{2011}

\maketitle

\label{firstpage}

\begin{abstract}
We investigate the properties (e.g. star formation rate, dust attentuation, stellar mass and metallicity) of a sample of infrared luminous galaxies at $z\sim1$ via near-IR spectroscopy with Subaru-FMOS. Our sample consists of {\it Herschel} SPIRE and {\it Spitzer} MIPS selected sources in the COSMOS field with photometric redshifts in the range $0.7<z_{\rm phot}<1.8$, which have been targeted in 2 pointings (0.5 sq.\,deg.) with FMOS. We find a modest success rate for emission line detections, with candidate H$\alpha$ emission lines detected for 57 of 168 SPIRE sources (34 per cent). By stacking the near-IR spectra we directly measure the mean Balmer decrement for the H$\alpha$ and H$\beta$ lines, finding a value of $\langle E(B-V)\rangle=0.51\pm0.27$ for $\langle L_{\rm IR}\rangle=10^{12}$\,$L_{\odot}$ sources at $\langle z \rangle=1.36$. By comparing star formation rates estimated from the IR and from the dust uncorrected H$\alpha$ line we find a strong relationship between dust attenuation and star formation rate. This relation is broadly consistent with that previously seen in star-forming galaxies at $z\sim0.1$. Finally, we investigate the metallicity via the N2 ratio, finding that $z\sim1$ IR-selected sources are indistinguishable from the local mass--metallicity relation. We also find a strong correlation between dust attentuation and metallicity, with the most metal-rich IR-sources experiencing the largest levels of dust attenuation.

\end{abstract}

\begin{keywords}
galaxies: evolution, submillimetre: galaxies
\end{keywords}

\section{Introduction}

Accurate measurements of the characteristic properties of galaxies; star formation rate (SFR), metallicity and stellar mass, are central to our understanding of their evolution. Significant progess in our ability to measure these properties in distant ($z>1$) galaxies has been made in the last two decades. Deep surveys with the {\it Hubble} space telescope have enabled the star formation rates of large numbers of galaxies up to $z\sim7$ to be measured (e.g.\ Madau et al.\ 1996; Bunker et al.\ 2004; Bouwens et al.\ 2006; Bouwens et al.\ 2009., McLure et al.\ 2010). Large scale optical and near-IR spectroscopic surveys have targetted key emission lines allowing the study of gas metallicity to $z<4$ (e.g.\ Tremonti et al.\ 2004; Erb et al.\ 2006; Mannucci et al.\ 2009; Zahid et al.\ 2011; Cresci et al.\ 2012). Finally, deep optical and near-IR photometric surveys have allowed an accurate assessment of the stellar mass contained in galaxies, and its build-up with redshift out to $z\sim5$ (e.g.\ Fontana et al.\ 2006; Ilbert et al.\ 2010; Caputi et al.\ 2011).

While these advances have re-shaped our understanding of galaxy formation and evolution, they typically rely observations in a single wavelength window i.e. UV, optical/near-IR, far-IR, etc. Meanwhile at low and intermediate redshifts it is becoming clear that large scale, multi-wavelength, studies of galaxies are needed to determine unbiased estimates of their properties. In the case of SFR estimates, where the corrections for dust attenuation tend to be large, comparisons of UV and IR SFR estimates at $z\sim0$ (Hao et al.\ 2011), $z\sim1$ (Buat et al.\ 2010) and $z\sim2$ (Reddy et al.\ 2011) show that widely used {\it in band} (i.e. in the same waveband as the SFR estimate) dust attenuation estimators (e.g. the UV continuum slope; Meurer et al.\ 1999) have large errors ($\delta\log_{10}SFR\sim0.3$ dex) and can have significant systemic biases for certain populations of galaxies (low SFR spirals, ULIRGs $10^{12}L_{\odot}$ and young starbursts). By comparison SFR estimators based on combinations of the far-IR, H$\alpha$ line or radio have much smaller errors ($\delta\log_{10}SFR\sim0.1$ dex; Kennicutt et al.\ 2009; Hao et al.\ 2011) and are universally valid.

Measurements of other galaxy properties also benefit from a multi-wavelength approach. Metallicity estimates from single tracers e.g. the $N2$ or $R_{32}$ methods (Pettini \& Pagel 2004) can disagree by up to $\Delta[\log_{10}(O/H)]=0.7$ dex (Kewley \& Ellison 2008). Stellar mass estimates obtained via the fitting the stellar population models require multi-band observations in the optical and near-IR to be reliable; omitting near-IR observations introduces an error of $\sim 0.1 dex$ to stellar mass estimates at $z\sim1$ (Pozzetti et al.\ 2007; Ilbert et al.\ 2010).

In order to put galaxy evolution at high$-z$ on a firm footing multi-wavelength observations at the same rest-frame wavelengths as our low-$z$ benchmarks (e.g. SDSS, {\it Spitzer}, {\it IRAS}) for a large number of high-$z$ galaxies is needed. This necessitates wide-field imaging and spectrocopy in the IR. 

Here we investigate the rest-frame optical-to-far IR properties of a sample of {\it Herschel}\footnote{\herschel\ is an ESA space observatory with science
instruments provided by Principal Investigator consortia.  It is open
for proposals for observing time from the worldwide astronomical
community.} (Pilbratt et al.\ 2010) sources which were targetted for near-IR spectroscopy with FMOS (Kimura et al.\ 2010). The key goal of this work is to determine the key galaxy properties (SFR, dust attenuation, stellar mass and metallicity) between a sample of high$-z$ ($0.8<z<1.7$), IR luminous ($>10^{11}\,$$L_{\odot}$) sources using the same tracers commonly used for low-$z$ samples. In this way we can be sure that our results are fully consistent (in terms of both calibration and selection effects) with those at low-$z$. The datasets used in this work are described in \S\ref{sec:data}, \S\ref{sec:fmosdetrate} presents the detection rate of H$\alpha$, \S\ref{sec:compspec} the aggregate near-IR spectral properties and \S\ref{sec:sfrcomp} a comparison of the star formation rates from the IR and H$\alpha$ line. In \S\ref{sec:smmet} we investigate the stellar mass and metallicity of our sample and, finally, \S\ref{sec:conc} summarises our conclusions. Throughout we assume a $\Lambda$CDM cosmology with $\Omega_{\Lambda}=0.7$, $\Omega_{\rm m}=0.3$ and $H_0=70$\,km\,s$^{-1}$\,Mpc$^{-1}$.

\section{Data}\label{sec:data}
\subsection{Pre-existing COSMOS data} 
The starting point for this work is the SPIRE observations of the COSMOS field (Scoville et al.\ 2007) taken as part of the {\it Herschel} Multi-tiered Extragalactic Survey (HerMES; Oliver et al.\ 2012). The SPIRE instrument, its in-orbit performance and
its scientific capabilities are described by Griffin et al.\ (2010);
its calibration methods and accuracy are outlined in Swinyard et al.\
(2010). Here we make use of SPIRE maps as described in Levenson et al.\ (2010). At the time of writing HerMES observations of COSMOS cover $\sim4.8$ sq.\,deg.\ to a 1$\sigma$ instrumental noise of $\sim$2\,mJy/Beam at the three SPIRE wavelengths of 250, 350 and 500$\,\mu$m.

As the SPIRE data offer an instrumental noise significantly lower than the confusion noise ($\sim6\,$mJy; Nyugen et al.\ 2010) we make use of prior source positions from higher angular resolution data to extract SPIRE photometry. The MIPS 24$\,\mu$m channel is the most obvious prior for SPIRE data as it offers a significant improvement in angular resolution (6 arcsec FWHM for MIPS 24$\,\mu$m vs. 18.6 arcsec FWHM for SPIRE 250$\,\mu$m) while also being able to account for $>80$ per cent of the flux at SPIRE wavelengths at the 24$\,\mu$m depths now available in a large fraction of HerMES fields (Bethermin et al.\ 2012; Oliver et al.\ 2012).

To construct our prior catalogue for SPIRE photometry we begin with the MIPS 24\,$\mu$m imaging from the {\it Spitzer} COSMOS survey (Le Floc'h et al.\ 2009). Here we make use of the publicly available imaging, performing source extraction via the {\sc starfinder} IDL package (Diolaiti et al.\ 2000).  The resulting catalogue covers $\sim2.1$ sq.\,deg.\ and has a typical 1$\sigma$ sensitivity of $\sigma=15\,\mu$Jy.

In order to provide the most accurate positional information, for both our SPIRE photometry and FMOS fibre positioning, we cross-match our 24$\,\mu$m catalogue to the publically available {\it HST} ACS $I_{\rm F814W}$-band  catalogue of Leauthaud et al.\ (2007). This catalogue covers 1.64 sq.\,deg.\ to a limiting magnitude of $I_{\rm F814W}<26.5$. In addition to improving the positional accuracy this matching helps eliminate spurious 24$\,\mu$m sources produced by artifacts in the image, in particular those located close to bright ($>1\,$mJy) sources. Of the 35,914 24$\,\mu$m sources located well within the ACS $i$-band coverage, 33,071 (92 per cent) have $I_{\rm F814W}$-band counterparts within 2 arcsec. 

SPIRE photometry is performed using the $I_{\rm F814W}$-band positions of the 24$\,\mu$m sources as a prior, following the algorithms described in Roseboom et al.\ (2010) and Roseboom et al.\ (2012). All $>3\sigma$ ($\sim60\,\mu$Jy) 24$\,\mu$m sources are considered as potential SPIRE counterparts. Using the residual map statistics we estimate that our prior-driven SPIRE catalogue reaches a typical point source sensitivity of $\sigma_{\rm tot}=$2.7, 3.5, and 3.2\,mJy at 250, 350 and 500$\,\mu$m, including the contribution from source confusion.

To complete our multi-wavelength COSMOS dataset we add multi-band optical/near-IR data and photometric redshifts from the catalogue of Ilbert et al. (2009). This catalogue is limited to $i_{\rm AB}^+<25$, and hence our HerMES-COSMOS sample is similarly restricted.

\subsection{FMOS observations and emission line measurements}\label{sec:fmoslines}
IR-selected sources were targeted in two pointings (0.5 sq.\,deg.) located within the COSMOS field with FMOS as part of the GTO program. The FMOS instrument (Kimura et al.\ 2008) consists of 400 1.2-arcsec diameter fibres which can be placed within a 30-arcmin diameter field of view. We used the low-resolution mode ($R\sim600$), allowing instantaneous coverage of both the J and H band ($0.9<\lambda<1.8\,\mu$m), with cross-beam switching i.e. two fibres for each target; one placed on the sky and one on the target, with the target/sky "switched" between them at regular intervals.

Potential targets for FMOS fibre allocation were selected from our HerMES-COSMOS parent catologue by requiring a photometric redshift in the range $0.65<z_{\rm phot}<1.75$ from the catalogue of Ilbert et al. (2009). This restriction was introduced to ensure that the H$\alpha$ line was likely within the FMOS wavelength coverage.

Fibre allocation preference was given to sources detected at both 24$\,\mu$m and 250$\,\mu$m ($>3\sigma_{\rm tot}$), followed by 24$\,\mu$m only sources. As well as science targets, a number (typically 2--4) of 2MASS (Skrutskie et al. 2006) selected stars were included in the observations for flux calibration purposes.
 
The first of our FMOS pointings was dedicated to solely HerMES-COSMOS targets, while for the second pointing (2010 November 24 and 25) we shared fibres with the evolSMURF project (Bunker et al., in prep). While the split between the samples was roughly 50-50, this was aided by the overlap between the samples (42 sources).

Table \ref{tab:fmosobs} details the exposure times, together with the number of 24 and 250$\,\mu$m detected (henceforth refered to as $24\cap250\,\mu$m), and 24$\,\mu$m only sources in each pointing. In total 241 fibres were allocated to IR-selected sources, with four sources appearing in both configurations, resulting in 237 unique targets (168 unique $24\cap250\,\mu$m targets). All data were reduced using the standard FMOS pipeline (Iwamuro et al., 2011). 

\begin{table}
\caption{Summary of FMOS observations}
\label{tab:fmosobs}
\begin{tabular}{lllll}
\hline
 & Date & T$_{\rm exp}$ & $N_{24\,\mu m}$ & $N_{24\cap250\,\mu m}$\\
\hline\hline
P1 & 2010 Nov. 22& 8$\,\times\,900$\,s & 136& 102\\
P2 & 2010 Nov. 24 \& 25 & 14$\,\times\,900$\,s & 105& 67  \\
\hline
\multicolumn{3}{c}{Total (Unique)} & 241 (237) & 169 (168) \\

\end{tabular}
\end{table}

Emission lines were identified in the 2D reduced frames, after flux calibration, via a semi-automated procedure.  At each pixel the line profile was fitted to the surrounding 9$\times$9 pixels. The pixel scale was 5\AA\/ in the spectral direction and 0.13 arcsec in the fibre direction. We only considered pixels in the wavelength ranges $1.1$--1.36$\,\mu$m and $1.42$--1.7$\,\mu$m. Pixels within 5\,\AA\/ of an OH line were excluded from consideration. The line profile was assumed to be Gaussian with {\sc FWHM}$=\lambda/600$\,\AA\/ in the spectral direction, and 6.9 pixels in the fibre direction. The noise was estimated by taking the variance of all illuminated pixels at that wavelength on the detector. Regions where the noise is exceptionally high ($>10\,\mu$Jy\,pixel$^{-1}$) were excluded. The local continuum was estimated by taking the median pixel value in a window of 20 pixels (200\,\AA), excluding the closest 7 pixels. For each fit the line signal-to-noise ratio (SNR), peak SNR and correlation coefficient between the line profile and 2D spectrum was measured. The line SNR ($\sigma_{\rm line}$) was calculated via Eqn.~\ref{eqn:snr},
\begin{equation}
\sigma_{\rm line}=\sum_{i} (d_{i}-c_{i})P_{i}/\sigma_{i}^2\left[\sum_i(P_{i}^2/\sigma_i^2)\right]^{-1/2}, 
\label{eqn:snr}
\end{equation}
where $d_i$ is the pixel intensity at position $i$, $c_i$ is the continuum at pixel $i$, $P_{i}$ the line profile at position $i$, and $\sigma_{i}$ the noise estimate at position $i$.

The peak SNR is defined as the ratio of the peak flux density, taken to be the mean flux density in a 3$\times$3 pixel window less the local continuum, to the standard deviation of the surrounding pixels in the spectrum.

Finally the correlation coefficient ($\rho_{\rm line}$) is calculated Eqn.~\ref{eqn:cc},
\begin{equation}
\rho_{\rm line}=\frac{\sum_{i}(d_{i}-\bar{d})(P_{i}-\bar{P})}{\sigma_{P}\sigma_d}
\label{eqn:cc}
\end{equation}
where $\sigma_{P}$ and $\sigma_{d}$ are the standard deviation of the line profile and data values, respectively.

Line fits which have $\sigma_{\rm line}>4$, $\sigma_{\rm peak}>2.5$ and $\rho_{\rm line}>20$ were considered as candidate emission lines. 

All candidate emission lines that have a wavelength within the range $(1+z_{\rm phot}-0.16)\times6563.4$\AA$<\lambda<(1+z_{\rm phot}+0.16)\times6563.4$\AA\/ were considered to be H$\alpha$. The window of $\delta z=0.16$ equates to $4\sigma_{\rm phot-z}$ for the typical photo-$z$ error quoted by Ilbert et al. (2009) at $z\sim1$. 

A total of 85 candidate H$\alpha$ emission lines were found from the sample of 237 unique 24$\,\mu$m targets. We assessed the reliability of our line identification technique in two ways. Firstly the line identification was repeated, but with the proposed redshifts (and hence wavelength search window) shifted. To ensure that the mock search windows are sufficiently far away from real lines, but still within the wavelength coverage of FMOS, sources at $z_{\rm phot}<1.1$ were given $z_{\rm mock}=z_{\rm phot}+0.16+\delta$, while those at $z_{\rm phot}>1.1$ were set to $z_{\rm mock}=z_{\rm phot}-0.15-\delta$, where $\delta$ is a random number between 0 and 0.15. As a result of this process seven lines were identified, giving an estimate of the false positive line detection rate of 7/85 or 8$\pm$3 per cent. No false lines were returned with $\sigma_{\rm line}>8$.

We compare the redshifts, as determined by the wavelength of our candidate H$\alpha$ lines, and the known spectroscopic redshifts. From our sample of 85 H$\alpha$ line emitters, 28 are found to have reliable (z$_{\rm qual}>3$) spectroscopic redshifts in the zCOSMOS bright ($i<22.5$) sample of Lilly et al.\ (2007). Of these, 27 (96 per cent) are found to be within $\delta z=0.01$ of our assumed H$\alpha$ redshift. No incorrect redshifts are found amongst the 11 sources with candidate H$\alpha$ lines at $\sigma_{\rm line}>8$ and a spectroscopic redshift from Lilly et al.\ (2007). While this result is encouraging, it is likely that the reliability of our line identification is a function of brightness, and the zCOSMOS bright sample is limited to $i_{\rm AB}^+<22.5$; roughly 54 per cent (46/85) of our candidate H$\alpha$ line emitters have $22.5<i_{\rm AB}^+<25$.

Via a similar process we can estimate the completeness of our line identification process. Assuming the random redshifts, $z_{\rm mock}$, described above, we inject mock emission lines into our data at a wavelength corresponding to $1+z_{\rm mock})\,6564.3$\AA\/. We estimate the completeness for the two pointings independantly. Table \ref{tab:lcomp} details the completeness (i.e. the ratio of sources detected to those injected) as a function of flux. No aperture correction is assumed, injected line fluxes are considered to be those contained within the 1.2 arcsec diameter fibre of FMOS. The completeness never reaches 100 per cent as lines at certain wavelengths will always be undetectable due to the gap in wavelength coverage from 1.36--1.42$\,\mu$m, as well as the masking of OH sky lines. In total 125 OH lines are suppressed which, combined with the gap due to atmospheric absorption, remove 24 per cent of the potential wavelength coverage. It can be seen that our identification process reaches this maximum level above a line flux of $f_{\rm H\alpha}\gs 4\times10^{-16}$ ergs\,cm$^{-2}$\,s$^{-1}$, while we are $\gs50$ per cent complete above a line flux of $f_{\rm H\alpha}\gs 1.5\times10^{-16}$ ergs\,cm$^{-2}$\,s$^{-1}$. The completeness estimates for the second pointing are marginally higher at $f_{\rm H\alpha}\ls 3\times10^{-16}$ ergs\,cm$^{-2}$\,s$^{-1}$ due to the increased exposure time (12600\,sec for the second pointing vs. 7200\,sec for the first). Note that these line fluxes consider only the flux contained within the 1.2 arcsec diameter fibres used on FMOS; no aperture correction has been applied yet.

\begin{table}
\caption{Completeness of line identification process as a function of flux. Completeness is estimated via injection of mock emission lines into the 2D spectra at random wavelengths/redshifts. Note that this is for flux contained within the 1.2 arcsec diameter fibre of FMOS; no aperture effects are considered.}
\label{tab:lcomp}
\begin{tabular}{lll}
Line flux & \multicolumn{2}{c}{Completeness}\\
 & P1 & P2\\
\hline
10$^{-16}$ ergs\,cm$^{-2}$\,s$^{-1}$. & per cent & per cent\\

\hline\hline
0.5 & 4.5 & 1.4\\
1. & 23.2 & 25.3\\
1.5 & 52.9 & 57.8\\
2. & 65.8 & 67.8\\
2.5 & 71.0 & 72.1\\
3. & 74.2 & 74.7\\
3.5 & 77.4 & 77.6\\
4. & 77.6 & 78.7\\

\hline
\end{tabular}
\end{table}

For each H$\alpha$ line emitter we attempt to measure the flux of the neighbouring [NII]\,6584 line. Line fluxes are estimated in the same manner as H$\alpha$, but with the central wavelength fixed at $(1+z)\,6584$\AA. Of the 85 H$\alpha$ line emitters; 33 also have [NII] at $SNR>3$.

Finally, both H$\alpha$ and [NII] line fluxes are corrected for the limited aperture and unknown stellar continuum. The aperture correction is determined via the ratio of $I_{\rm F814W}$ flux within the FMOS fibre (1.2 arcsec diameter) to that within the Kron radius. The typical aperture correction is $\sim2$--3. For the continuum correction, the near-IR spectra are not deep enough to detect the continuum emission near the H$\alpha$ line. Thus, continuum emission at the wavelength of the line is estimated, and removed, using the available broadband optical and near-IR imaging from the COSMOS survey (Capak et al.\ 2007, McCracken et al.\ 2010). For sources where the line lies at $\lambda<1.4\,\mu$m we use the $J$ band magnitude, and the $[z^+-J]$ colour to estimate the continuum flux. For those lines where $\lambda>1.4\,\mu$m the $K$ band magnitude and $[J-K]$ colour are used. In all cases we assume that the H$\alpha$ line is coincident with stellar absorption of equivalent width $EW=4.4$ (Moustakas \& Kennicutt 2006).

Appendix \ref{ap:bigtable} contains the line fluxes and photometric properties for our 85 H$\alpha$ line emitters. While we list the full catalogue of H$\alpha$ line detections resulting from our observations, in the following analysis we consider only the 57 H$\alpha$ line emitters which have robust detections at 250$\,\mu$m. This restriction is implemented as 24$\,\mu$m alone is not a good tracer of total IR luminosity at these redshifts (Elbaz et al.\ 2010). Enforcing 250$\,\mu$m detections also helps to minimise contamination by AGN activity (Hatziminaoglou et al.\ 2010).

\subsection{SDSS comparison sample}
To establish a low redshift baseline for our $z\sim1$ measurements we identify a sample of {\it Spitzer} 160$\,\mu$m selected sources from the SWIRE survey (Lonsdale et al.\ 2003) in the Lockman Hole field with SDSS DR6 (Adelman-McCarthy et al.\ 2008) spectroscopy. Sources are required to be brighter than $50\,$mJy at 160$\,\mu$m and to lie at $z<0.4$. The 160$\,\mu$m photometry is complemented with SWIRE MIPS 24, 70$\,\mu$m data, and SPIRE photometry at 250, 350 and 500$\,\mu$m from HerMES. H$\alpha$ line fluxes are calculated using the {\sc GANDALF} package (Sarzi et al.\ 2006), with aperture corrections based on the ratio of the SDSS Petrosian-to-fibre $r$-magnitudes. We retain sources with a peak line flux density to noise ratio of greater than three. The typical limiting H$\alpha$ line flux (3$\sigma$) is $5.8\times10^{16}$\,erg\,cm$^{-2}$\,s$^{-1}$. Our final low-$z$ sample consists of 156 160$\,\mu$m-detected sources with reliable H$\alpha$ line flux measurements, at a mean redshift of $z=0.1$.

\section{Results}\label{sec:results}
\subsection{H$\alpha$ detection rate for IR-selected sources}\label{sec:fmosdetrate}
\begin{figure}

\includegraphics[scale=0.33]{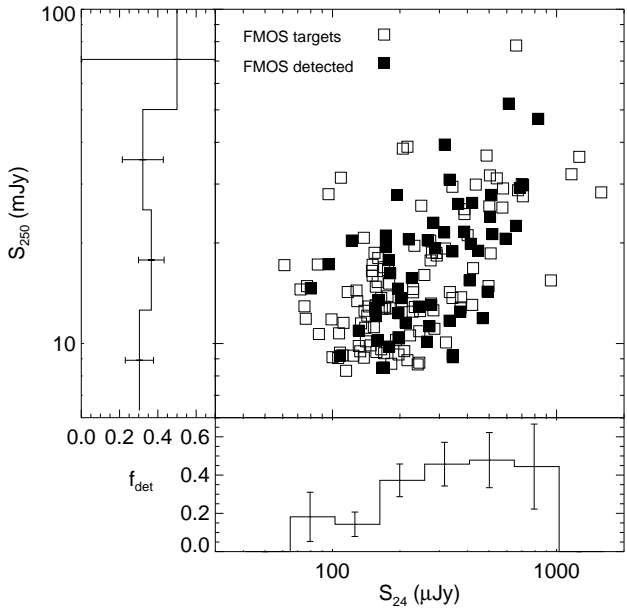}
\caption{H$\alpha$ line detection rate of FMOS targeted sources as a function of 24$\,\mu$m and 250$\,\mu$m flux density. The detection rate is relatively insensitive to 250$\,\mu$m flux density, while a modest gain is seen for  sources with $S_{24\,\mu m}>200\,\mu$Jy. }
\label{fig:fmosdet} 
\end{figure}

The raw H$\alpha$ detection rate for IR-sources (24$\cap250\,\mu$m) is 57/168 (34$\pm4$ per cent). In Fig.~\ref{fig:fmosdet} we present the H$\alpha$ detection rate as a function of 250$\,\mu$m and 24$\,\mu$m flux densities. While the H$\alpha$ detection rate appears to be insensitive to 250$\,\mu$m flux density, a modest increase in the detection rate is seen for sources with $S_{24\,\mu m}>200\,\mu$Jy.

\begin{figure}

\includegraphics[scale=0.35]{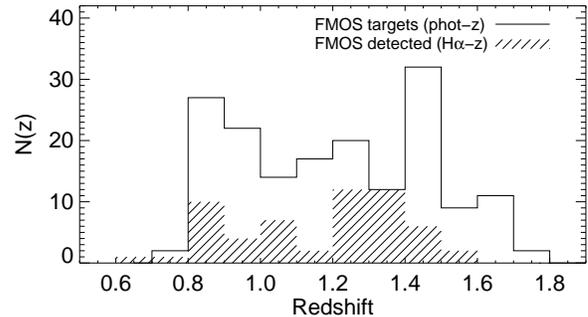}
\caption{Redshift distribution of FMOS targeted sources and sources with H$\alpha$ line detections. For FMOS targets we use photometric redshifts from Ilbert et al.\ (2009). For detected sources the redshift implied by the location of the H$\alpha$ line is used.}
\label{fig:fmoszhist} 
\end{figure}

Fig.~\ref{fig:fmoszhist} shows the redshift distribution of IR-sources targeted with FMOS, and those with H$\alpha$ line detections. The contrast between the two distributions highlights the visibility of H$\alpha$ with FMOS as a function of redshift. Prior to FMOS observations, only photometric redshifts were available for the vast majority of targets. Thus while targets were photo$-z$ selected to be in the redshift range where H$\alpha$ is visible, some fraction of sources were expected to have H$\alpha$ fall at wavelengths outside the FMOS wavelength range.

Despite this the observed detection rate compares well with what would be predicted given the known completeness (\S\ref{sec:fmoslines}). For each detected source we calculate the completeness at that line flux, $c_{l}$, by interpolating the values in Table \ref{tab:lcomp}. If the H$\alpha$ line fluxes of the undetected sources are distributed similarly to the detected ones (i.e. the bulk of the incompleteness is due to OH sky lines) then the sum $\sum_{l}1/c_{l}$ should be equal to the number of targetted objects. For our 24$\cap250\,\mu$m sample $\sum_{l}1/c_{l}=177$; close to the actual number of sources targetted (168).

We can compare our detection rate for IR-selected sources with FMOS to optically selected sources with VLT VIMOS. For the VVDS-DEEP survey (Le F\'{e}vre et al.\ 2005),  the detection rate for $i$-selected sources with similar magnitudes (median for our sample is $i_{\rm AB}^+\sim23.5$) and redshifts ($z\sim1.2$) to our sample is $\sim 70$ per cent (for 4.5\,hr exposures; Ilbert et al.\ 2005). While our FMOS detection rate is almost one-half of this, IR-sources have quite a low areal density and cannot make the most of the large multiplex of VIMOS. For our parent sample of 24 \& 250$\,\mu$m detected, $i_{\rm AB}^+<25$ and $0.65<z_{\rm phot}<1.75$ sources the areal density is $\sim1500$ per sq.\,deg. Using these numbers as a guide, the expected number of redshifts recovered in a single 2\,hr FMOS pointing (0.36$\times200$ fibres equals 72 sources), is comparable to the number expected from a single 4.5\,hr VIMOS pointing (0.7$\times91$ targets equals 64). Thus for {\it Herschel} selected targets FMOS is a competitive facility for redshift recovery in the range $0.7<z<1.8$.

\subsection{Composite spectrum for H$\alpha$ detected sources}\label{sec:compspec}
While our near-IR spectra are of sufficient quality to robustly measure fluxes for bright emission lines, very little additional information can be extracted from the individual spectra. This is partly due to the low SNR, but also because of the large number of OH sky lines in the near-IR, which significantly limit the wavelength coverage. However, using the spectral coverage unaffected by OH sky lines from each of our H$\alpha$-detected sources we can build a composite spectrum across a reasonably wide and continuous range of rest-frame wavelengths.

\begin{figure}
\includegraphics[scale=0.58]{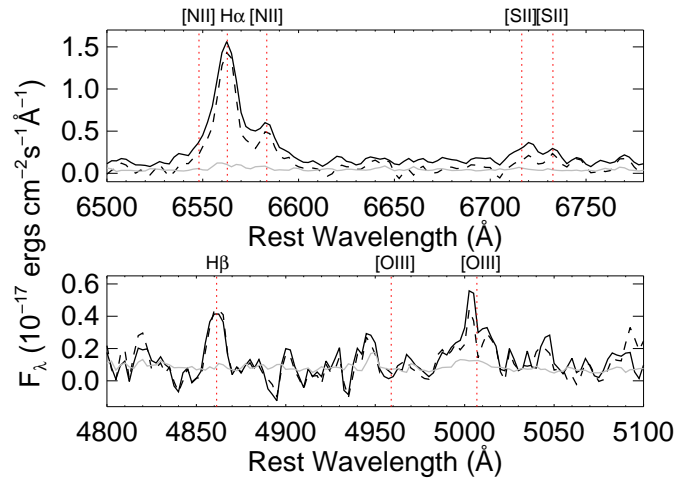}
\caption{Composite rest-frame spectrum for the 57 $24\cap250\,\mu$m sources with robust line detections in the region of H$\alpha$ (top) and H$\beta$ (bottom). The dashed line is the composite spectrum for only those 24 sources which are visible at the rest frame wavelength of H$\beta$. The grey solid line represents the 1$\sigma$ variance in the composite. Well-known spectral features are marked. Encouragingly, our composite spectrum recovers several other spectral lines, such as [NII]\,6584, the blended doublet of [SII]\,6716 and [SII]\,6731, and weak signatures of [OIII]\,5008 and H$\beta$. }
\label{fig:fmosstack}
\end{figure}

Fig.~\ref{fig:fmosstack} shows the composite rest-frame spectrum in the region of H$\alpha$ and H$\beta$ produced from our sample of 57 sources. The composite spectrum is produced by first adopting a grid in rest-frame wavelength and calculating the contribution of each observed near-IR spectrum via a Gaussian kernel. The {\sc FWHM} of the Gaussian kernel is taken to be the spectral resolution, $\lambda/600$\,\AA. Pixels within 10\,\AA\ of an OH-suppressed line are excluded, as are those which lie outside the wavelength range $1.1$--$1.36\,\mu$m or $1.42$--$1.7\,\mu$m. The variance in the composite spectrum is estimated via jack-knife resampling of the sources contributing to each wavelength bin. The absolute flux calibration of the composite is corrected by comparing the continuum level in the composite to the mean broadband photometry in the $J$ and $K$ bands.

Several well-known spectral lines can be seen in addition to H$\alpha$: [NII]\,6584, although this is blended with H$\alpha$; a blend of the [SII] doublet at 6716 and 6731\,\AA\/; and weak signatures of [OIII]\,5008 and H$\beta$. 

The line ratio H$\alpha$/[NII]\,6584 compared to the line ratio [OIII]\,5008/H$\beta$ is often used as diagnostic of AGN activity (Baldwin, Phillips \& Terlevich 1981; BPT). Measuring the strengths of these lines from Fig.~\ref{fig:fmosstack} we determine $f_{\rm [NII]\,6584}/f_{\rm H\alpha}=0.24\pm0.07$, while $f_{\rm [OIII]\,5008}/f_{\rm H\beta}=1.3\pm0.6$. Comparing these ratios with the BPT diagnostic plots suggests that, on average, the primary origin of emission lines, and by proxy the IR-luminosity, in our sample is star formation. 

The standard approach to estimating the dust attenuation in star-forming galaxies is to use the flux ratios of Balmer lines. Under typical conditions (i.e. T$_{\rm e}\sim10^4$\,K), and in the absence of dust, the line ratio $f_{\rm H\alpha}/f_{\rm H\beta}=2.86$ (Osterbrock 1989). Observed differences in this ratio, also known as the Balmer decrement, can be attributed to differential dust attenuation at the rest-frame wavelengths of the Balmer lines. 

Using our composite spectrum we estimate the aggregate value of the Balmer decrement for SPIRE sources at $z\gs1$. The H$\beta$ line is visible in the FMOS wavelength coverage at $1.26<z<2.5$, excluding sky lines. To account for this we build a second composite spectrum, using only those sources which have `clean' (i.e. no overlapping sky lines) FMOS wavelength coverage at the wavelength of both H$\beta$ and H$\alpha$; only 24 sources satisfy this criterion. The composite spectrum from these sources in the region of H$\alpha$ and H$\beta$ is shown in Fig.~\ref{fig:fmosstack}. The H$\beta$ observable sources have a mean redshift of $\langle z\rangle=1.36$ and a mean IR luminosity of $\langle L_{\rm IR} \rangle=10^{45.5}$\,ergs\,s$^{-1}$ (10$^{12}\,$$L_{\odot}$).  We measure line fluxes of $\langle f_{\rm H\alpha}\rangle$=6.8$\pm0.6\times10^{-16}$\,ergs\,cm$^{-2}$s$^{-1}$ and $\langle f_{\rm H\beta} \rangle$=1.3$\pm0.4\times10^{-16}$\,ergs\,cm$^{-2}$s$^{-1}$, after correcting both lines for stellar absorption (EW$_{\rm H\alpha}=4.4$ and EW$_{\rm H\beta}$=2.8\AA; Moustakas \& Kennicutt 2006) and applying an aperture correction of 2.8 (the mean value for these 24 sources; see \S\ref{sec:sfrcomp}). This gives a Balmer decrement of $\langle R\rangle=f_{\rm H\alpha}/f_{\rm H\beta}=5.2\pm1.6$, resulting in a dust attenuation of $\langle E(B-V) \rangle=\log_{10}(R/2.86)/0.4[k(\lambda_{\rm H\alpha})-k(\lambda_{\rm H\beta})]=0.51\pm0.26$, where $k(\lambda_{\rm H\alpha})=4.596$ and $k(\lambda_{\rm H\beta})=3.325$ (Calzetti et al.\ 2000). This is equivalent to $A_v=2.1$ mags, similar to that found for local IR-luminous galaxies (e.g. Hopkins et al.\ 2001, Wijesinghe et al.\ 2011).

\subsection{Relationship between H$\alpha$ and IR star formation rate estimates}\label{sec:sfrcomp}
In the absence of AGN activity or strongly non-solar metallicity, differences between the H$\alpha$ estimated SFR (SFR$_{\rm H\alpha}$), and the best estimate of the total SFR (SFR$_{\rm tot}$) can be attributed to the effect of dust attenuation. Thus the ratio SFR$_{\rm tot}$/SFR$_{\rm H\alpha}$ can be used as an estimator of the level of dust attenuation (e.g. Hopkins et al.\ 2001; Kewley et al.\ 2002). In order to calculate SFR$_{\rm tot}$ and SFR$_{\rm H\alpha}$ the following steps were taken.

IR luminosities are calculated by fitting template SEDs to the 24$\,\mu$m and SPIRE data, then integrating the best fit template in the range $8$--$1000\,\mu$m. IR template SEDs are taken from Rieke et al.\ (2009). Both IR and H$\alpha$ luminosities are converted into SFR via the relations presented in Kennicutt (1998), assuming a Chabrier (2003) IMF, i.e;
\begin{equation}
{\rm SFR}_{\rm IR}=2.61\times10^{-44}L_{\rm IR}~{\rm (erg\,s^{-1})},
\end{equation}
and

\begin{equation}
{\rm SFR}_{\rm H\alpha}=4.61\times10^{-42}L_{\rm H\alpha}~{\rm (erg\,s^{-1})}
\end{equation}
Finally, ${\rm SFR}_{\rm tot}$ is calculated by adding together the IR and H$\alpha$ star formation rates, i.e.
\[{\rm SFR}_{\rm tot}={\rm SFR}_{\rm IR}+{\rm SFR}_{\rm H\alpha}.\]

\begin{figure}
\includegraphics[scale=0.5]{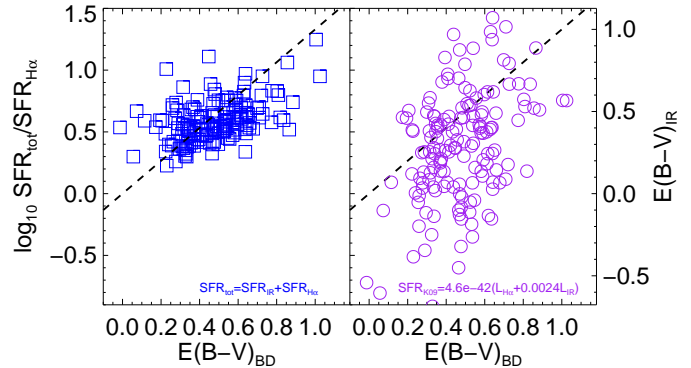}
\caption{Left panel: Comparison of dust attenuation estimates from the Balmer decrement ($E(B-V)_{\rm BD}$) and from the ratio of SFR$_{\rm tot}$/SFR$_{\rm H\alpha}$ ($E(B-V)_{\rm IR}$) for SDSS galaxies. The dashed line represents $E(B-V)_{\rm IR}= E(B-V)_{\rm BD}$. The two estimates are excellent agreement, with a RMS difference of $0.17$. Right panel: Comparison of dust attenuation estimates from the Balmer decrement ($E(B-V)_{\rm BD}$) and from the ratio of SFR$_{\rm K09}$/SFR$_{\rm H\alpha}$, where SFR$_{\rm K09}$ is the combined H$\alpha$+IR SFR estimator presented in Kennicutt et al.\ 2009 (K09). Again, the dashed line represents $E(B-V)_{\rm IR}= E(B-V)_{\rm BD}$ It can be seen that the K09 SFR calibration gives much poorer agreement with the Balmer decrement estimated attenuations (RMS$=0.27$).}
\label{fig:hsdssebv}
\end{figure}

While other combined H$\alpha$+IR estimators of SFR exist (e.g. Kennicutt et al.\ 2009), these may not give good results for the class of IR-luminous galaxies (nor {\it Herschel} derived IR luminosities). We can confirm our assumption that SFR$_{\rm tot}$/SFR$_{\rm H\alpha}$ is a good tracer of dust attenuation by comparing to estimates of the attenuation from the Balmer decrement for our SDSS sample. All 156 of our SDSS comparison sample have reliable ($>3\sigma$) H$\beta$ line flux estimates. We produce estimates of the dust attenuation independant of the far-IR measurements using the Balmer decrement. Fig.~\ref{fig:hsdssebv} compares the dust attenuation estimated from the Balmer decrement ($E(B-V)_{\rm BD}$) to the infered value assuming SFR$_{\rm tot}$/SFR$_{\rm H\alpha}$ is a good estimate of $A_{\rm H\alpha}$.

\begin{figure}
\includegraphics[scale=0.43]{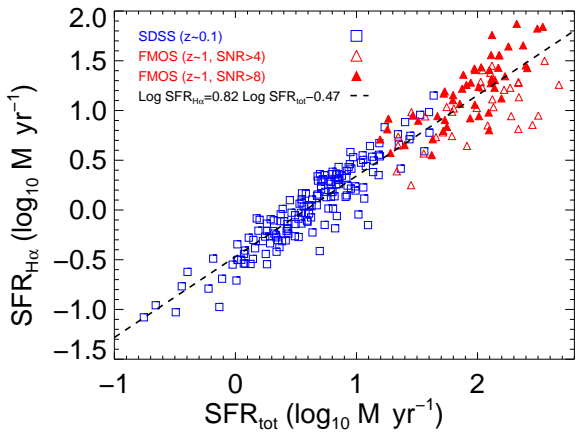}
\caption{Comparison of H$\alpha$ to  total (H$\alpha$ + Far-IR) SFRs, assuming the conversions of Kennicutt (1998) with a Chabrier (2003) IMF. FMOS detected sources (red triangles) and the low-$z$ sample of 160$\,\mu$m-selected SDSS galaxies (blue squares) are shown. The dashed line shows the best-fit log-linear relation to the data.}

\label{fig:fmosll} 
\end{figure}

Fig.~\ref{fig:fmosll} shows the relationship between SFR$_{\rm H\alpha}$ and SFR$_{\rm tot}$ for our sample of 57 sources detected at 24$\,\mu$m, 250$\,\mu$m and H$\alpha$.  Fitting this observed correlation with a log-linear function results in the best fit;

\[\log_{10}{\rm SFR}_{\rm H\alpha}=(0.82\pm0.08)\log_{10}{\rm SFR}_{\rm IR}+0.47\pm0.11 ~({\rm M_{\odot}\,yr^{-1}})\]

Using the Calzetti et al.\ 2000 model for the variation of dust attenuation with wavelength we can use this result to convert the ratio SFR$_{\rm tot}$/SFR$_{\rm H\alpha}$ to $E(B-V)$. Taking our best fit correlation between these values we derive the following relationship between SFR and dust attenuation;
\begin{equation}
E(B-V)_{\rm IR}= (0.135\pm0.06) \log_{10}{\rm SFR}_{\rm tot} + 0.35\pm0.08 ~({\rm M_{\odot}\,yr^{-1}})
\label{eqn:ebvsfr}
\end{equation}

\begin{figure}
\includegraphics[scale=0.26]{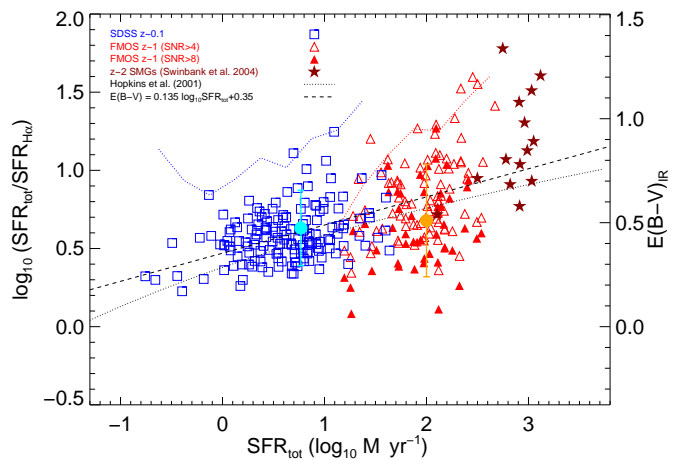}
\caption{Ratio of IR to dust uncorrected H$\alpha$-based SFR vs. IR-based star formation rate (SFR). The equivalent $E(B-V)$ is also given, assuming, a Calzetti et al. (2000) attenuation curve. FMOS detected sources (red triangles) and the low-$z$ sample of 160$\,\mu$m-selected SDSS galaxies (blue squares) are shown. The dashed line shows the local relation between SFR and $E(B-V)$ from Hopkins et al.\ (2001). The $E(B-V)$ determined from the Balmer decrement of our FMOS composite spectrum (Fig.~\ref{fig:fmosstack}), and mean value for SDSS spectra are shown as cyan and orange circles, respectively. Also shown are the positions of $z\sim2$ SMGs with H$\alpha$ line measurements from Swinbank et al.\ (2004). Dotted lines show the effect of the H$\alpha$ detection limit (5$\sigma$) at the mean redshift for sources with that SFR$_{\rm tot}$ for the SDSS (blue) and FMOS (red) samples.}
\label{fig:fmossfr} 
\end{figure}

Fig.~\ref{fig:fmossfr} presents dust uncorrected SFR$_{\rm tot}$/SFR$_{\rm H\alpha}$ vs. SFR$_{\rm tot}$. We also show in Fig.~\ref{fig:fmossfr} our best fit relationship between $E(B-V)$ and SFR$_{\rm tot}$, and the empirical relationship for low-$z$ {\it IRAS} galaxies as derived by Hopkins et al.\ (2001). These two relations show reasonable agreement, with the Hopkins et al. (2001) relation slightly below our simple log-linear fit.

Alternative estimates of the mean (and standard deviation) of dust attenuation measured from the Balmer decrement of individual SDSS spectra, and the aggregate FMOS value from our composite spectrum (\S\ref{sec:compspec}) are also shown in Fig.~\ref{fig:fmossfr}. Encouragingly the direct estimates of the mean dust attenuation are in good agreement with that inferred from SFR$_{\rm IR}$/SFR$_{\rm H\alpha}$, and both the dust attenuation-SFR relations.

In Fig.~\ref{fig:fmossfr} are also shown {\it unlensed} 850$\,\mu$m selected sources (SMGs) at $z\gs2$, with H$\alpha$ line measurements from Swinbank et al.\ (2004). Here we make use of the SFR$({\rm H\alpha})$ and L(FIR) quantities given in Table 2 of Swinbank et al.\ (2004), converting SFR$({\rm H\alpha})$ to the Chabrier (2003) IMF used here and calculating SFR$_{\rm IR}$ from L(FIR) using the equation given above. SFR$({\rm H\alpha})$ as quoted by Swinbank et al.\ (2004) includes corrections for slit-loss and so should be compatible with the values we derive from FMOS and SDSS data. Interestingly the $z\gs2$ SMGs appear slightly above both the Hopkins et al.\ (2001) and our best-fit dust attenuation -- SFR relation, suggesting that SMGs experience enhanced attenuation. However it is worth noting that the L(FIR) estimates for the SMGs come from pre-{\it Herschel} submm-radio estimates and hence may be overestimated (see Magnelli et al.\ 2012). Future studies with FMOS in this SFR range, as well as a re-assessment of the SMG population with {\it Herschel} photometry will allow this trend to be confirmed.

The modest H$\alpha$ detection limits achievable with FMOS mean we will not recover IR-sources which have very large SFR$_{\rm tot}$/SFR$_{\rm H\alpha}$. To quantify this we calculate the typical maximum observable limit of SFR$_{\rm tot}$/SFR$_{\rm H\alpha}$ as a function of SFR$_{\rm tot}$ for both the SDSS and FMOS samples. For the SDSS sample we assume a detection limit of f$_{\rm H\alpha}=5.8\times10^{16}$\ergs, while for the FMOS sample we assume a detection limit of f$_{\rm H\alpha}=1\times10^{16}$\ergs. In both cases these limits include a correction for the mean loss due to the limited aperture of the fibres (1.2 arcsec diameter for FMOS, 3 arcsec diameter for SDSS). While the maximum limit for the SDSS sample is significantly higher ($\sim0.2$ dex) than the observed values of SFR$_{\rm tot}$/SFR$_{\rm H\alpha}$, the limits for the FMOS dataset appear quite close to the observed data points.

Given $\sim65$ per cent of our parent sample is undetected in H$\alpha$ a potential explanation for this large incompleteness is a significant population of sources with SFR$_{\rm tot}$/SFR$_{\rm H\alpha}$ above these selection limits. Hence the observed consistency with the Hopkins et al.\ (2001), and our Eqn.~\ref{eqn:ebvsfr} may be a result of a bias towards low SFR$_{\rm tot}$/SFR$_{\rm H\alpha}$.

 While we cannot rule out the existence of large SFR$_{\rm tot}$/SFR$_{\rm H\alpha}$ sources (as we cannot detect them), we can estimate the {\it observed} completeness for our parent sample assuming our best-fit to the $E(B-V)$--SFR$_{\rm tot}$ relation is a good description for the whole population. For each source in our parent sample of 168 $24\cap250\,\mu$m sources we first estimate $L_{\rm IR}$, assuming the photo-$z$ from Ilbert et al. (2009) and the SED fitting process described above. The H$\alpha$ line flux is then predicted from $L_{\rm IR}$ (assuming SFR$_{\rm IR}=$SFR$_{\rm tot}$) using the best fit log-linear relation from Fig.~\ref{fig:fmossfr} (with 0.5 dex of intrinsic scatter) and the mean loss due to the fibre aperture (2.8). Applying the completeness curves from Table~\ref{tab:lcomp} we would expect a completeness of 35 per cent, in good agreement with the observed completeness of 34$\pm4$ per cent, and consistent with a similar assessment of the completeness presented in \S\ref{sec:fmosdetrate}.

If a large fraction of our parent IR-selected sample had dust attenuation levels (i.e. SFR$_{\rm tot}$/SFR$_{\rm H\alpha}$) significantly above that predicted by Eqn.~\ref{eqn:ebvsfr} we would expect much lower completeness in our FMOS observations than achieved. Given the good agreement between the observed and expected completeness we conclude that Eqn.~\ref{eqn:ebvsfr} must hold for the bulk of IR-luminous sources at $z\sim1$.
\subsection{The mass--metallicity relation for IR-galaxies}\label{sec:smmet}
A significant fraction (28/57) of our sample have robust (SNR$>3$) measurements of the [NII]\,6584 line. This allows us to investigate the gas phase metallicity (12+$\log_{10}$\,O/H) of our IR-sources via the ratio of the [NII]\,6584 to the H$\alpha$ line (N2 method; Kewley \& Dopita 2002; Pettini \& Pagel 2004). For each source with a robust [NII] measurement we estimate the metallicity via Eqn. 1 of Pettini \& Pagel (2004; PP04);\[12+\log_{10}\,{\rm O/H} = 8.9+0.57\times \log_{10}\,f_{\rm [NII]}/f_{\rm H\alpha}\]

The existence of a relationship between stellar mass and metallicity has been confirmed across a wide range in redshift ($0<z<3$; Lequeux et al.\ 1979; Tremonti et al.\ 2004; Erb et al.\ 2006; Maiolino et al.\ 2008; Zahid et al.\ 2011), with a steady trend towards lower metallicity for a given stellar mass with increasing redshift. To investigate the mass--metallicity relation for our $z\sim1$ IR-sources we combine our metallicity estimates with stellar masses as derived by Wang et al.\ (2012). Stellar mass estimates from Wang et al.\ (2012) are calculated by finding the best-fit stellar population model to the observed multi-wavelength photometry using the {\sc Le Phare} software (Arnouts et al.\ 2002; Ilbert et al.\ 2006), combined with stellar population synthesis models from Bruzual \& Charlot (2003) and assuming a Chabrier (2003) IMF. 

\begin{figure*}
\includegraphics[scale=0.45]{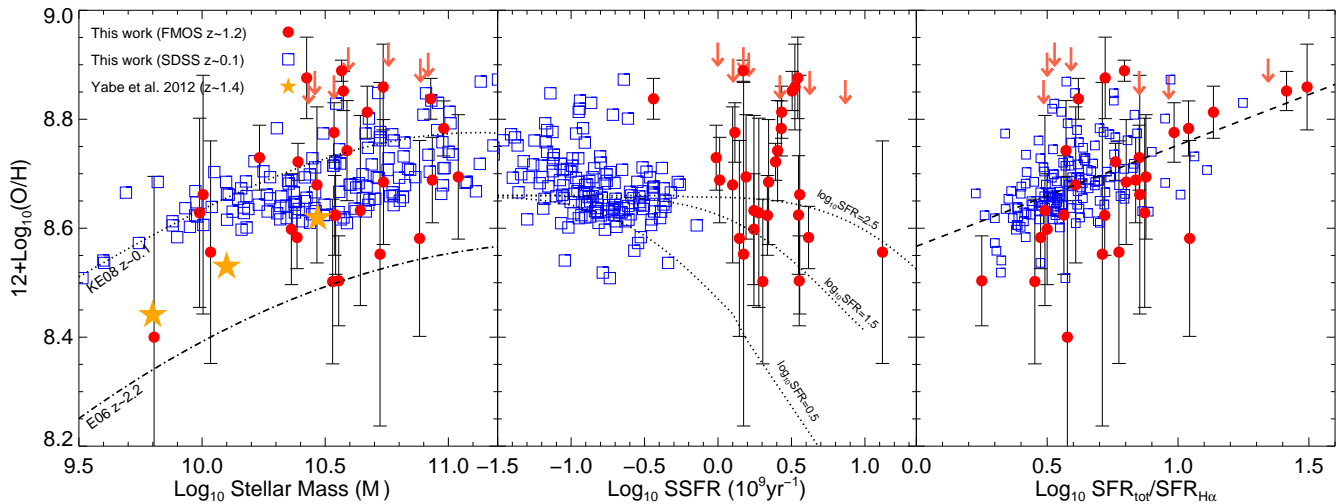}
\caption{Relationship between stellar mass (left panel), specific SSFR (middle panel), dust attenuation (right panel) vs. gas phase metallicity as seen in our IR-selected sample. All sources with [NII] line flux measurements with SNR$>3$ are shown, while limits are shown for those sources without robust [NII] line measures. Sources from our {\it Herschel}-FMOS sample at $z\sim1.2$ are shown as red dots, while the $\sim0.1$ {\it Herschel}-SDSS samples are blue squares. Previous estimates of the stellar mass -- metallicity relation at $z\sim0.1$ (Kewley \& Ellison (2008); KE08) and $z\sim2.2$ (Erb et al.\ 2006; E06) are shown as dotted and dot-dashed lines in the left panel, respectively. Results from a previous FMOS study for star-forming galaxies at $z\sim1.4$ (Yabe et al.\ 2012) are shown as orange stars. Where necessary, comparison samples have been converted to use the Pettini \& Pagel (2004) calibration of the N2 metallicity tracer (KE08) and a Chabrier (2003) IMF. In the middle panel the position of the ``fundamental metallicity relation'' of Mannucci et al.\ (2010) is shown for a range of SFRs representative of our sample as the dotted lines. In the right panel the best-fit log-linear relation between metallicity and dust attenuation (as traced by the ratio SFR$_{\rm tot}$/SFR$_{\rm H\alpha}$) is shown as a dashed line. }
\label{fig:smmet} 
\end{figure*}

The left panel of Fig.~\ref{fig:smmet} compares the stellar mass to the metallicity for both our sample of $z\sim1$ {\it Herschel}-FMOS and comparison sample of $z\sim0.1$ SDSS sources. A tentative trend of metallicity with stellar mass can be seen in both samples, although with significant scatter. No discernable evolution is seen in the metallicity between the $z\sim1$ and $z\sim0.1$ IR-selected samples.

Also shown in the left panel of Fig.~\ref{fig:smmet} are studies using stacking of near-IR spectra by Erb et al.\ (2006) with Keck NIRSPEC at $z\sim2.2$ and Yabe et al.\ (2012) with FMOS at $z\sim1.4$ (We convert the Yabe et al.\ 2012 stellar masses to a Chabrier 2003 IMF for consistency). Given the similar redshift range between our {\it Herschel}-FMOS sample and the Yabe et al.\ (2012) study the difference in the metallcities is surprising, although likely due to a combination of the different mass ranges probed, the stacking nature of the Yabe et al.\ (2012) result and our bias towards dusty galaxies.

In the middle panel of Fig.~\ref{fig:smmet} we show the relationship between specific SFR (SSFR;SFR/M$_{\star}$) and metallicity. Over the range in SSFR well-sampled by our {\it Herschel}-FMOS and SDSS samples we see no discernable trends, although there are hints of decreasing metallicity with increasing SSFR above SSFR$=0.5$\,Gyr$^{-1}$. It has been proposed that the metallicity is a function of both SFR and stellar mass, with a ``fundamental mass relation'' (FMR) linking the three parameters (Mannucci et al.\ 2010). In Fig.~\ref{fig:smmet} we show the fit to the FMR from Mannucci et al.\ (2010) for values of SFR representative of our study. The metallcities of both our {\it Herschel}-FMOS and SDSS are consistent with the prediction from the FMR.

Finally, we consider the relationship between dust attenuation and metallicity. In the right panel of Fig.~\ref{fig:smmet} is shown the metallicity as a function of dust attenuation (as traced by the ratio SFR$_{\rm tot}$/SFR$_{\rm H\alpha}$). A clear trend between dust attenuation and metallicity is seen in both the {\it Herschel}-FMOS and SDSS samples, with the most metal-rich galaxies experiencing the largest levels of attenuation. The best fit log-linear relation between attenuation and metallicity is found to be:
\[12+\log_{10}\,{\rm O/H}=0.19\,\log_{10}{\rm SFR}_{\rm tot}/{\rm SFR}_{\rm H\alpha}+8.57.\]

Converting to $E(B-V)$ via the Calzetti et al.\ (2000) attenuation curve we find Eqn.~\ref{eqn:metebv},
\begin{equation}
12+\log_{10}\,{\rm O/H}=0.24\,E(B-V)+8.57
\label{eqn:metebv}
\end{equation}

This correlation is somewhat expected; the dust grains responsible for attenuating starlight (and nebular line emission) are synthesised from metals in the ISM. Similar studies, using the IR to UV luminosity ratios as a proxy for dust attentuation, have also found a correlation between metallicity and attenuation for nearby galaxies (Heckman et al.\ 1998) and $z\sim2$ (Reddy et al.\ 2010). Given the correlation between metallicity with attenuation the high metallicity of our {\it Herschel}-FMOS sample (when compared to Yabe et al.\ 2012 at $z\sim1.4$) is to be expected as our IR-selected sample must be biased towards the most obscured sources. 
 
\section{Discussion and Conclusions}\label{sec:conc}
We have investigated the properties of $z\sim1$ IR luminous galaxies by performing near-IR spectroscopy of a sample of {\it Spitzer} and {\it Herschel} selected sources with FMOS. Candidate emission lines were identified in the 2D reduced FMOS frames via a semi-automated procedure. Via comparison with known spectroscopic redshifts, and direct testing of the line detection algorithm, we estimate that our H$\alpha$ line sample is $\gs90$ per cent reliable. Our scientific conclusions are;
\begin{itemize}
\item Robust detections of H$\alpha$ were found for 57 of 168 24$\cap250\,\mu$m-sources, resulting in a detection rate of 34$\pm4$ per cent. This detection rate is consistent with the expected incompleteness measured by simulating the line detection process. For sparse targets such are {\it Herschel}-selected sources, FMOS is a competitive redshift recovery instrument with equivalent optical MOS instruments on other 8m class telescope (e.g. VIMOS on VLT).

\item The mean dust attenuation, estimated via the Balmer decrement for the H$\alpha$ and H$\beta$ emission lines for a composite spectrum of H$\alpha$ detected sources, is $E(B-V)=0.51\pm0.27$ for $L_{\rm IR}=10^{12}$\,$L_{\odot}$ sources at $\langle z\rangle=1.36$. 

\item  Good agreement was found between the dust attenuation estimated from the Balmer decrement and that inferred from the ratio of H$\alpha$-estimated to best estimate of the total star formation rate. Using SFR$_{\rm tot}$/SFR$_{\rm H\alpha}$ as an indicator of attenuation in the H$\alpha$ line we derive a relationship between dust attenuation and star formation rate of $E(B-V)=(0.135\pm0.06)\log_{10}$SFR$_{\rm tot}+0.35\pm0.08$. These results are broadly consistent with the relationship between star formation rate and dust attenuation seen both in low-$z$ star forming galaxies (e.g. Hopkins et al.\ 2001).

\item The gas phase metallicity relation was investigated for the subset of {\it Herschel}-FMOS, and SDSS, sources with robust measurements of the [NII]\,6584 line. For IR-selected sources with M$_{\star}\sim10^{10.5}$M$_{\odot}$ the typical metallicity is found not to evolve between $z\sim0.1$ to $z\sim1.2$. No discernable trend with specfic SFR is seen, but a strong correlation between metallicity and dust attentuation is seen, described by a best-fit log-linear relation; $12+\log_{10}\,{\rm O/H}=0.24\,E(B-V)+8.57$.

\end{itemize}

\section*{Acknowledgements}
We thank the anonymous referee for suggestions which greatly enhanced this work.

IGR, SJO and LW acknowledge support from the Science and Technology Facilities Council [grant number ST/F002858/1] and [grant number ST/I000976/1]

JSD acknowledges the support of the Royal Society via a Wolfson Research Merit award, and the support of the European Research Coucil via the award of an Advanced Grant.\\

%

Based on zCOSMOS observations carried out using the Very Large Telescope at the ESO Paranal Observatory under Programme ID: LP175.A-0839.\\

Funding for the SDSS and SDSS-II has been provided by the Alfred P. Sloan Foundation, the Participating Institutions, the National Science Foundation, the U.S. Department of Energy, the National Aeronautics and Space Administration, the Japanese Monbukagakusho, the Max Planck Society, and the Higher Education Funding Council for England. The SDSS Web Site is http://www.sdss.org/.

The SDSS is managed by the Astrophysical Research Consortium for the Participating Institutions. The Participating Institutions are the American Museum of Natural History, Astrophysical Institute Potsdam, University of Basel, University of Cambridge, Case Western Reserve University, University of Chicago, Drexel University, Fermilab, the Institute for Advanced Study, the Japan Participation Group, Johns Hopkins University, the Joint Institute for Nuclear Astrophysics, the Kavli Institute for Particle Astrophysics and Cosmology, the Korean Scientist Group, the Chinese Academy of Sciences (LAMOST), Los Alamos National Laboratory, the Max-Planck-Institute for Astronomy (MPIA), the Max-Planck-Institute for Astrophysics (MPA), New Mexico State University, Ohio State University, University of Pittsburgh, University of Portsmouth, Princeton University, the United States Naval Observatory, and the University of Washington.

SPIRE has been developed by a consortium of institutes led by Cardiff
Univ.\ (UK) and including Univ.\ Lethbridge (Canada); NAOC (China);
CEA, LAM (France); IFSI, Univ.\ Padua (Italy); IAC (Spain); Stockholm
Observatory (Sweden); Imperial College London, RAL, UCL-MSSL, UK ATC,
Univ.\ Sussex (UK); Caltech, JPL, NHSC, Univ.\ Colorado (USA). This
development has been supported by national funding agencies: CSA
(Canada); NAOC (China); CEA, CNES, CNRS (France); ASI (Italy); MCINN
(Spain); SNSB (Sweden); STFC and UKSA (UK); and NASA (USA). \\

FMOS was funded jointly by STFC and the Japanese Monbukagakusho, and we gratefully acknowledge the support of the staff at the Subaru Telescope throughout the instrument commissioning phase.\\

The data presented in this paper will be released through the {\em Herschel} Database in Marseille HeDaM ({hedam.oamp.fr/HerMES})\\

\newpage
\appendix
\section{Details of the HerMES-FMOS sample}\label{ap:bigtable}
\begin{scriptsize}
\onecolumn
\begin{longtable}{cccccccccc}
\caption{Details of {\it Herschel} 250$\,\mu$m and {\it Spitzer} 24$\,\mu$m-selected sources with reliable H$\alpha$ line detections.}\\
\label{tab:bigtable}
RA$^{a}$ & Dec. & Redshift$^b$ & $I_{\rm F814W}$ & S$_{24\,\mu m}$ & S$_{250\,\mu m}^c$ & S$_{350\,\mu m}^c$ & S$_{500\,\mu m}^c$ & f$_{\rm H\alpha}^d$ & f$_{\rm [NII]}^e$ \\
\hline
deg. & deg. &  & AB mag & mJy & mJy & mJy & mJy & 10$^{16}$\,erg\,cm$^{-2}$\,s$^{-1}$ & 10$^{16}$\,erg\,cm$^{-2}$\,s$^{-1}$\\
\hline\hline
\endfirsthead
\caption{(continued)}\\
RA$^a$ & Dec. & Redshift$^b$ & $I_{\rm F814W}$ & S$_{24\,\mu m}$ & S$_{250\,\mu m}^c$ & S$_{350\,\mu m}^c$ & S$_{500\,\mu m}^c$ & f$_{\rm H\alpha}^d$ & f$_{\rm [NII]}^e$ \\
\hline
deg. & deg. &  & AB & mJy & mJy & mJy & mJy & 10$^{16}$\,erg\,cm$^{-2}$\,s$^{-1}$ & 10$^{16}$\,erg\,cm$^{-2}$\,s$^{-1}$\\
\hline\hline
\endhead
\hline
\endfoot
\multicolumn{10}{l}{$^{a}$ Position from $I_{\rm F814W}$ catalogue.}\\
\multicolumn{10}{l}{$^{b}$ Redshift from detected H$\alpha$ line.}\\
\multicolumn{10}{l}{$^{c}$ Limits are $3\sigma$, including confusion noise.}\\
\multicolumn{10}{l}{$^{d}$ All values corrected for the limited aperture (1.2 arcsec diameter) of the FMOS fibre.}\\
\multicolumn{10}{l}{$^{e}$ All values corrected for the limited aperture (1.2 arcsec diameter) of the FMOS fibre. Limits are $3\sigma$, no value given in cases where the [NII]\,6584 line falls on a OH sky line.}\\

\endlastfoot
  149.94205 &    2.32077 & 1.03 & 21.94$\pm0.01$ & 0.20$\pm0.01$ &  9.80 $\pm  2.73$ & $< 12.22$ & $< 15.98$ &  4.32$\pm 0.30$ & $< 3.95$\\
 149.96689 &    2.44185 & 0.90 & 22.44$\pm0.02$ & 0.08$\pm0.01$ & $<  8.20$ & $< 13.15$ & $<  9.28$ &  2.84$\pm 0.32$ & $< 5.78$\\
 149.97292 &    2.49010 & 1.36 & 23.64$\pm0.03$ & 0.43$\pm0.01$ & 15.44 $\pm  2.73$ & 19.04 $\pm  3.22$ & $< 15.41$ &  1.52$\pm 0.41$ & $< 6.17$\\
 149.95179 &    2.48627 & 1.03 & 22.00$\pm0.01$ & 0.28$\pm0.01$ & 10.13 $\pm  2.74$ & 11.98 $\pm  3.38$ & $< 14.81$ &  2.33$\pm 0.17$ & $< 2.62$\\
 149.99931 &    2.45197 & 1.51 & 23.13$\pm0.03$ & 0.33$\pm0.01$ & 21.61 $\pm  2.73$ & 15.34 $\pm  3.61$ & 18.14 $\pm  4.87$ &  2.96$\pm 0.40$ & $< 3.05$\\
 150.03484 &    2.26353 & 0.90 & 21.46$\pm0.01$ & 0.25$\pm0.01$ & 11.26 $\pm  2.73$ & $< 26.43$ & $< 33.24$ &  5.44$\pm 0.37$ & $< 6.56$\\
 150.04760 &    2.62123 & 1.38 & 22.73$\pm0.02$ & 0.23$\pm0.01$ & 20.49 $\pm  2.73$ & 13.12 $\pm  3.25$ & $< 17.00$ &  3.20$\pm 0.27$ & $< 2.19$\\
 150.02476 &    2.35211 & 0.93 & 21.76$\pm0.01$ & 0.15$\pm0.01$ & $<  8.20$ & $< 13.09$ & $< 35.23$ &  4.55$\pm 0.38$ & $< 5.37$\\
 150.03650 &    2.31730 & 1.46 & 22.36$\pm0.02$ & 0.44$\pm0.01$ & 26.35 $\pm  2.73$ & 25.45 $\pm  3.48$ & $< 17.85$ & 11.43$\pm 0.33$ & $< 4.66$\\
 150.02440 &    2.44388 & 1.03 & 21.72$\pm0.02$ & 0.39$\pm0.01$ & $< 41.18$ & $< 32.57$ & $< 25.69$ & 10.69$\pm 0.56$ & $< 7.19$\\
 150.04820 &    2.46775 & 1.17 & 23.33$\pm0.03$ & 0.34$\pm0.01$ & 11.71 $\pm  2.73$ & $<  9.31$ & $< 14.30$ &  7.69$\pm 0.34$ &  2.57$\pm 0.78$\\
 150.03637 &    2.49426 & 1.03 & 21.86$\pm0.01$ & 0.41$\pm0.01$ &  9.23 $\pm  2.74$ & $<  9.59$ & $< 15.54$ &  6.86$\pm 0.28$ & $< 3.75$\\
 150.08336 &    2.53619 & 1.42 & 23.70$\pm0.04$ & 0.30$\pm0.01$ & 22.99 $\pm  2.73$ & 25.93 $\pm  4.79$ & 33.71 $\pm  9.86$ &  7.97$\pm 0.28$ & $< 3.39$\\
 150.05928 &    2.51858 & 1.03 & 22.10$\pm0.01$ & 0.38$\pm0.01$ & 12.46 $\pm  2.74$ & $<  9.61$ & $< 16.43$ &  1.92$\pm 0.13$ & $< 2.17$\\
 150.05386 &    2.58972 & 0.70 & 18.89$\pm0.00$ & 3.45$\pm0.01$ & $<  8.19$ & $< 10.02$ & $< 14.47$ & 68.67$\pm 0.31$ & 47.99$\pm 1.39$\\
 150.10975 &    2.60274 & 0.98 & 21.80$\pm0.01$ & 0.32$\pm0.01$ & $<  8.21$ & $< 33.17$ & $<  9.27$ & 10.12$\pm 0.19$ & --\\
 150.10019 &    2.48157 & 0.89 & 22.11$\pm0.02$ & 0.08$\pm0.01$ & $<  8.23$ & $<  9.27$ & $< 14.32$ &  4.88$\pm 0.48$ & $< 5.28$\\
 150.13340 &    2.26201 & 0.75 & 21.63$\pm0.01$ & 0.08$\pm0.01$ & $<  8.19$ & $<  9.94$ & $< 14.36$ &  3.31$\pm 0.32$ & $< 4.71$\\
 150.14563 &    2.29341 & 0.88 & 22.27$\pm0.02$ & 0.10$\pm0.01$ & $<  8.20$ & $<  9.29$ & $<  4.77$ &  3.77$\pm 0.37$ & $< 4.26$\\
 150.13445 &    2.61448 & 0.89 & 22.44$\pm0.01$ & 0.15$\pm0.01$ & $<  8.22$ & $< 16.66$ & $< 16.33$ &  2.93$\pm 0.16$ & $< 1.78$\\
 150.16162 &    2.69151 & 1.48 & 23.51$\pm0.03$ & 0.52$\pm0.01$ & 21.25 $\pm  2.73$ & 24.13 $\pm  3.95$ & 17.66 $\pm  5.36$ &  5.05$\pm 0.61$ & --\\
 150.15563 &    2.67708 & 1.04 & 22.28$\pm0.02$ & 0.21$\pm0.01$ & 10.36 $\pm  2.74$ & $< 29.40$ & $< 32.03$ &  3.14$\pm 0.15$ & $< 2.70$\\
 150.12472 &    2.66871 & 1.28 & 22.56$\pm0.01$ & 0.19$\pm0.01$ & 21.05 $\pm  2.73$ & $< 12.30$ & $< 14.92$ &  5.82$\pm 0.12$ &  1.43$\pm 0.45$\\
 150.15229 &    2.21933 & 0.92 & 21.86$\pm0.01$ & 0.36$\pm0.01$ & $<  8.21$ & $< 12.19$ & $<  7.61$ &  3.60$\pm 0.19$ & $< 2.96$\\
 150.21805 &    2.52182 & 1.18 & 21.72$\pm0.01$ & 0.30$\pm0.01$ & 13.05 $\pm  2.74$ & 22.89 $\pm  3.56$ & $< 16.66$ &  5.98$\pm 0.34$ & $< 2.59$\\
 150.21025 &    2.56547 & 1.40 & 22.29$\pm0.02$ & 0.34$\pm0.01$ & 30.92 $\pm  2.73$ & 22.22 $\pm  3.69$ & $< 36.15$ &  5.34$\pm 0.15$ &  2.24$\pm 0.66$\\
 150.25980 &    2.29235 & 0.99 & 22.42$\pm0.01$ & 0.14$\pm0.01$ & $<  8.20$ & $< 16.07$ & $< 18.66$ &  3.48$\pm 0.27$ & --\\
 150.22858 &    2.31620 & 0.90 & 21.25$\pm0.01$ & 0.30$\pm0.01$ & $<  8.22$ & $<  9.56$ & $< 17.02$ &  8.68$\pm 0.17$ & $< 2.55$\\
 150.22203 &    2.62003 & 0.69 & 21.50$\pm0.01$ & 0.65$\pm0.01$ & 27.91 $\pm  2.73$ & 22.13 $\pm  7.22$ & $< 21.83$ &  9.14$\pm 0.48$ & $< 9.70$\\
 150.24813 &    2.39912 & 0.68 & 21.03$\pm0.01$ & 0.15$\pm0.01$ & $<  2.73$ & $<  3.35$ & $< 17.64$ &  9.01$\pm 0.50$ & $< 6.14$\\
 150.29832 &    2.46967 & 0.85 & 21.75$\pm0.01$ & 0.18$\pm0.01$ & $<  8.19$ & $< 10.17$ & $< 25.53$ &  2.91$\pm 0.20$ & $< 3.79$\\
 150.29408 &    2.51691 & 0.84 & 21.89$\pm0.01$ & 0.16$\pm0.01$ & $<  8.21$ & $< 18.92$ & $< 12.42$ &  3.02$\pm 0.30$ & $< 0.84$\\
 150.25672 &    2.48480 & 1.24 & 23.13$\pm0.03$ & 0.19$\pm0.01$ & 17.77 $\pm  2.74$ & 12.31 $\pm  3.66$ & $< 30.69$ &  4.84$\pm 0.64$ &  4.32$\pm 0.66$\\
 150.29414 &    2.57633 & 0.78 & 22.40$\pm0.02$ & 0.21$\pm0.01$ & $<  8.21$ & $< 11.11$ & $< 19.23$ &  1.98$\pm 0.20$ & $< 5.92$\\
 150.31964 &    2.61813 & 1.00 & 21.59$\pm0.01$ & 0.24$\pm0.01$ & $< 30.62$ & $< 31.33$ & $< 24.63$ &  5.03$\pm 0.36$ & --\\
 150.27952 &    2.59666 & 1.49 & 23.03$\pm0.02$ & 0.17$\pm0.01$ & 13.49 $\pm  2.73$ & 13.22 $\pm  3.15$ & $< 15.18$ &  3.81$\pm 0.47$ & $< 2.73$\\
 150.34435 &    2.73311 & 0.85 & 21.56$\pm0.01$ & 0.26$\pm0.02$ & $<  8.19$ & 15.09 $\pm  3.95$ & $< 48.57$ &  2.33$\pm 0.12$ &  1.10$\pm 0.14$\\
 150.32742 &    2.68368 & 0.96 & 21.62$\pm0.01$ & 0.45$\pm0.01$ & $<  8.21$ & $< 18.05$ & $< 15.67$ &  9.08$\pm 0.22$ &  3.84$\pm 0.28$\\
 150.34866 &    2.45537 & 1.02 & 22.12$\pm0.02$ & 0.22$\pm0.01$ & 11.54 $\pm  2.74$ & $<  9.68$ & $< 21.07$ &  5.77$\pm 0.25$ &  1.67$\pm 0.27$\\
 150.31731 &    2.50830 & 0.98 & 22.31$\pm0.01$ & 0.09$\pm0.01$ & 14.61 $\pm  2.74$ & 12.64 $\pm  3.38$ & $< 33.01$ &  2.73$\pm 0.29$ &  1.01$\pm 0.30$\\
 150.36745 &    2.55928 & 0.91 & 22.20$\pm0.02$ & 0.13$\pm0.01$ & $<  8.19$ & $<  8.07$ & $< 16.49$ &  2.94$\pm 0.26$ & $< 4.69$\\
 150.35415 &    2.51916 & 0.96 & 21.32$\pm0.01$ & 0.75$\pm0.01$ & 29.87 $\pm  2.74$ & $< 23.46$ & $< 31.47$ &  6.18$\pm 0.50$ &  1.63$\pm 0.30$\\
 150.35350 &    2.55854 & 1.04 & 21.63$\pm0.01$ & 0.17$\pm0.01$ & 10.25 $\pm  2.73$ & $< 13.46$ & $< 14.98$ &  4.17$\pm 0.24$ &  3.08$\pm 0.30$\\
 150.34722 &    2.57026 & 0.92 & 21.48$\pm0.01$ & 0.30$\pm0.01$ & $<  8.19$ & $< 20.10$ & $< 22.18$ &  3.01$\pm 0.19$ & $< 3.12$\\
 150.36110 &    2.60817 & 1.30 & 22.89$\pm0.02$ & 0.11$\pm0.01$ & $<  8.19$ & $< 20.03$ & $< 24.99$ &  3.96$\pm 0.21$ & $< 0.58$\\
 150.33580 &    2.64975 & 1.33 & 22.58$\pm0.02$ & 0.12$\pm0.01$ & 20.33 $\pm  2.73$ & 23.03 $\pm  3.78$ & $< 14.14$ &  3.66$\pm 0.21$ &  1.49$\pm 0.20$\\
 150.34937 &    2.36957 & 1.57 & 23.49$\pm0.04$ & 0.13$\pm0.02$ & $<  8.23$ & $< 18.35$ & $< 15.06$ &  7.85$\pm 0.70$ &  2.70$\pm 0.64$\\
 150.39351 &    2.44529 & 0.92 & 21.67$\pm0.01$ & 0.28$\pm0.01$ & $<  8.19$ & $< 11.84$ & $< 15.06$ &  7.62$\pm 0.25$ &  1.18$\pm 0.26$\\
 150.37070 &    2.49822 & 0.82 & 20.52$\pm0.01$ & 0.43$\pm0.01$ & 19.88 $\pm  2.73$ & 11.93 $\pm  3.54$ & $< 17.47$ &  8.12$\pm 0.28$ &  7.45$\pm 0.66$\\
 150.41783 &    2.55859 & 1.21 & 22.13$\pm0.01$ & 0.76$\pm0.01$ & 47.01 $\pm  2.74$ & 34.07 $\pm  3.15$ & 19.10 $\pm  4.88$ &  2.47$\pm 0.32$ &  1.97$\pm 0.30$\\
 150.42288 &    2.58332 & 0.82 & 20.43$\pm0.01$ & 0.69$\pm0.01$ & 29.20 $\pm  2.74$ & $< 28.48$ & $< 15.74$ & 30.07$\pm 0.65$ &  3.95$\pm 1.03$\\
 150.42188 &    2.32331 & 0.83 & 23.16$\pm0.02$ & 0.61$\pm0.01$ & 20.59 $\pm  2.74$ & $<  9.86$ & $< 18.94$ & 16.98$\pm 1.98$ & --\\
 150.40063 &    2.33453 & 1.21 & 22.24$\pm0.01$ & 0.36$\pm0.01$ & 26.17 $\pm  2.74$ & 21.99 $\pm  3.20$ & $< 15.97$ &  5.93$\pm 0.47$ &  2.43$\pm 0.43$\\
 150.42068 &    2.62304 & 1.29 & 23.56$\pm0.04$ & 0.22$\pm0.01$ & 14.60 $\pm  2.73$ & $< 15.63$ & $< 16.45$ &  4.58$\pm 0.21$ & --\\
 150.44572 &    2.76095 & 1.35 & 23.11$\pm0.03$ & 0.18$\pm0.01$ & 12.08 $\pm  2.74$ & $< 10.95$ & $< 21.60$ &  2.02$\pm 0.18$ & $< 0.74$\\
 150.49513 &    2.37789 & 0.89 & 21.62$\pm0.01$ & 0.62$\pm0.01$ & 52.22 $\pm  2.73$ & 40.12 $\pm  3.56$ & 22.55 $\pm  6.23$ &  3.88$\pm 0.25$ & $< 1.12$\\
 150.48572 &    2.71969 & 0.89 & 21.95$\pm0.01$ & 0.58$\pm0.01$ & 27.84 $\pm  2.72$ & 17.44 $\pm  3.12$ & $< 15.71$ &  7.59$\pm 0.33$ & $< 1.65$\\
 150.51527 &    2.56268 & 1.27 & 23.40$\pm0.03$ & 0.20$\pm0.01$ & 12.75 $\pm  2.74$ & $< 51.16$ & $< 46.81$ &  6.56$\pm 0.64$ &  3.43$\pm 0.42$\\
 150.51142 &    2.57740 & 0.79 & 21.35$\pm0.01$ & 0.14$\pm0.01$ & 10.93 $\pm  2.74$ & 13.80 $\pm  3.40$ & $< 16.00$ &  2.88$\pm 0.24$ &  0.91$\pm 0.21$\\
 150.50345 &    2.65037 & 1.29 & 23.02$\pm0.03$ & 0.22$\pm0.02$ & 13.67 $\pm  2.73$ & 13.00 $\pm  3.09$ & 15.19 $\pm  4.77$ &  1.52$\pm 0.11$ & --\\
 150.55036 &    2.73248 & 0.85 & 20.93$\pm0.01$ & 0.52$\pm0.01$ & 14.30 $\pm  2.74$ & $< 10.32$ & $< 18.21$ & 10.68$\pm 0.49$ &  3.41$\pm 0.32$\\
 150.56193 &    2.47613 & 0.82 & 20.86$\pm0.01$ & 0.48$\pm0.01$ & 11.91 $\pm  2.73$ & $< 19.96$ & $< 16.95$ &  9.00$\pm 0.28$ &  4.25$\pm 0.34$\\
 150.54374 &    2.49755 & 0.88 & 21.73$\pm0.01$ & 0.15$\pm0.02$ & $<  8.21$ & $< 28.80$ & $< 33.36$ &  5.60$\pm 0.48$ &  3.26$\pm 0.61$\\
 150.59733 &    2.61799 & 1.41 & 21.75$\pm0.00$ & 0.28$\pm0.01$ & 20.36 $\pm  2.73$ & $< 45.00$ & 18.81 $\pm  5.39$ &  1.72$\pm 0.15$ & --\\
 150.56845 &    2.64125 & 1.26 & 23.55$\pm0.03$ & 0.19$\pm0.01$ & 16.23 $\pm  2.75$ & $< 10.33$ & $< 11.99$ &  8.51$\pm 0.46$ & --\\
 150.59931 &    2.38340 & 0.89 & 21.78$\pm0.01$ & 0.37$\pm0.01$ &  9.10 $\pm  2.73$ & $< 12.41$ & $< 15.17$ &  3.58$\pm 0.22$ &  2.07$\pm 0.25$\\
 150.59031 &    2.53405 & 1.50 & 22.60$\pm0.01$ & 0.18$\pm0.01$ &  8.46 $\pm  2.74$ & 10.75 $\pm  3.54$ & $< 18.56$ &  3.20$\pm 0.25$ & --\\
 150.62185 &    2.55274 & 0.88 & 21.45$\pm0.01$ & 0.46$\pm0.01$ & 18.95 $\pm  2.74$ & 15.05 $\pm  3.73$ & 16.41 $\pm  5.12$ &  6.22$\pm 0.66$ &  2.49$\pm 0.43$\\
 150.63049 &    2.55622 & 1.29 & 22.29$\pm0.02$ & 0.12$\pm0.01$ &  9.19 $\pm  2.74$ & $< 45.35$ & $< 52.71$ &  5.96$\pm 0.25$ & --\\
 150.63901 &    2.61703 & 1.33 & 22.80$\pm0.02$ & 0.22$\pm0.01$ & 12.32 $\pm  2.72$ & $<  9.65$ & $< 15.44$ & 15.27$\pm 0.41$ &  3.06$\pm 0.32$\\
 150.62021 &    2.62575 & 1.32 & 21.14$\pm0.01$ & 0.34$\pm0.01$ & 21.49 $\pm  2.74$ & $< 11.06$ & $< 18.35$ &  3.00$\pm 0.20$ &  1.99$\pm 0.19$\\
 150.62429 &    2.72533 & 1.21 & 22.73$\pm0.02$ & 0.17$\pm0.01$ &  8.48 $\pm  2.73$ & $< 10.81$ & $< 11.59$ &  6.74$\pm 0.26$ &  1.34$\pm 0.24$\\
 150.66639 &    2.44691 & 0.87 & 21.51$\pm0.01$ & 0.32$\pm0.01$ & $<  8.23$ & $< 16.02$ & $< 16.12$ &  5.26$\pm 0.33$ & --\\
 150.63611 &    2.74441 & 1.38 & 23.63$\pm0.04$ & 0.35$\pm0.01$ & 39.39 $\pm  2.73$ & $< 34.02$ & $< 29.14$ &  5.12$\pm 0.31$ &  3.12$\pm 0.33$\\
 150.66376 &    2.49340 & 1.34 & 22.26$\pm0.02$ & 0.27$\pm0.01$ & 12.90 $\pm  2.73$ & $< 11.30$ & $< 15.90$ &  3.48$\pm 0.24$ & --\\
 150.67631 &    2.52723 & 1.38 & 24.10$\pm0.05$ & 0.69$\pm0.01$ & 22.50 $\pm  2.73$ & 20.77 $\pm  3.66$ & $< 21.77$ &  3.43$\pm 0.41$ & $< 0.76$\\
 150.68868 &    2.61353 & 0.96 & 22.15$\pm0.02$ & 0.36$\pm0.01$ & 18.86 $\pm  2.73$ & $< 11.03$ & $< 12.07$ &  3.97$\pm 0.21$ & --\\
 150.66590 &    2.64705 & 1.40 & 22.45$\pm0.02$ & 0.18$\pm0.01$ & 13.48 $\pm  2.74$ & $<  9.29$ & $<  4.78$ &  6.61$\pm 0.16$ &  1.81$\pm 0.20$\\
 150.70877 &    2.52972 & 0.98 & 21.54$\pm0.01$ & 0.49$\pm0.01$ & 23.98 $\pm  2.73$ & 11.34 $\pm  3.28$ & $< 16.55$ &  3.34$\pm 0.19$ &  2.58$\pm 0.26$\\
 150.70577 &    2.56658 & 0.90 & 22.51$\pm0.02$ & 0.07$\pm0.01$ & $< 20.36$ & $< 11.73$ & $< 17.71$ &  1.02$\pm 0.11$ & $< 0.52$\\
 150.72242 &    2.61250 & 1.20 & 22.76$\pm0.03$ & 0.23$\pm0.01$ & 15.70 $\pm  2.74$ & $< 12.59$ & $< 17.81$ &  4.58$\pm 0.46$ &  1.10$\pm 0.31$\\
 150.31065 &    2.31922 & 1.43 & 25.53$\pm0.13$ & 0.17$\pm0.01$ & $<  8.21$ & $< 18.13$ & $< 19.54$ &  3.28$\pm 0.26$ & $< 2.85$\\
 150.37685 &    2.45953 & 1.30 & 25.53$\pm0.14$ & 0.36$\pm0.01$ & 19.29 $\pm  2.74$ & 22.76 $\pm  4.10$ & $< 19.70$ &  4.51$\pm 0.38$ & $< 2.60$\\
 150.46438 &    2.63802 & 1.46 & 24.59$\pm0.10$ & 0.09$\pm0.01$ & 17.27 $\pm  2.74$ & 25.30 $\pm  3.15$ & 26.53 $\pm  4.91$ &  1.80$\pm 0.32$ & $< 1.50$\\
 150.68825 &    2.45364 & 1.38 & 25.95$\pm0.15$ & 0.18$\pm0.01$ & 19.43 $\pm  2.73$ & 16.53 $\pm  3.66$ & $< 17.36$ &  6.78$\pm 0.45$ &  2.20$\pm 0.27$\\
\hline
\end{longtable}
\end{scriptsize}

\label{lastpage}


\begin{thebibliography}{99}
\bibitem[Adelman-McCarthy et al.(2008)]{2008ApJS..175..297A} 
Adelman-McCarthy, J.~K., et al.\ 2008, \apjs, 175, 297 
\bibitem[Arnouts et al.(2002)]{2002MNRAS.329..355A} Arnouts, S., 
Moscardini, L., Vanzella, E., et al.\ 2002, \mnras, 329, 355
\bibitem[Baldwin et al.(1981)]{1981PASP...93....5B} Baldwin, J.~A., 
Phillips, M.~M., \& Terlevich, R.\ 1981, \pasp, 93, 5 
\bibitem[Bouwens et al.(2009)]{2009ApJ...705..936B} Bouwens, R.~J., 
Illingworth, G.~D., Franx, M., et al.\ 2009, \apj, 705, 936
\bibitem[Bruzual 
\& Charlot(2003)]{2003MNRAS.344.1000B} Bruzual, G., \& Charlot, S.\ 2003, \mnras, 344, 1000
\bibitem[Capak et al.(2007)]{2007ApJS..172...99C} Capak, P., Aussel, H., 
Ajiki, M., et al.\ 2007, \apjs, 172, 99 
\bibitem[Calzetti et al.(2000)]{2000ApJ...533..682C} Calzetti, D., Armus, 
L., Bohlin, R.~C., Kinney, A.~L., Koornneef, J., 
\& Storchi-Bergmann, T.\ 2000, \apj, 533, 682
\bibitem[Cresci et al.(2012)]{2012MNRAS.421..262C} Cresci, G., Mannucci, 
F., Sommariva, V., et al.\ 2012, \mnras, 421, 262
\bibitem[Choi et al.(2006)]{2006ApJ...637..227C} Choi, P.~I., et al.\ 2006, 
\apj, 637, 227 
\bibitem[Diolaiti et al.(2000)]{2000SPIE.4007..879D} Diolaiti, E., 
Bendinelli, O., Bonaccini, D., Close, L.~M., Currie, D.~G., 
\& Parmeggiani, G.\ 2000, \procspie, 4007, 879 
\bibitem[Elbaz et 
al.(2010)]{2010A&A...518L..29E} Elbaz, D., et al.\ 2010, \aap, 518, L29 
\bibitem[Erb et al.(2006)]{2006ApJ...644..813E} Erb, D.~K., Shapley, A.~E., 
Pettini, M., et al.\ 2006, \apj, 644, 813 
\bibitem[Garn et al.(2010)]{2010MNRAS.402.2017G} Garn, T., et al.\ 2010, 
\mnras, 402, 2017
\bibitem[Griffin et 
al.(2010)]{2010A&A...518L...3G} Griffin, M.~J., et al.\ 2010, \aap, 518, L3
\bibitem[Hatziminaoglou et 
al.(2010)]{2010A&A...518L..33H} Hatziminaoglou, E., et al.\ 2010, \aap, 518, L33 
\bibitem[Heckman et al.(1998)]{1998ApJ...503..646H} Heckman, T.~M., Robert, 
C., Leitherer, C., Garnett, D.~R., 
\& van der Rydt, F.\ 1998, \apj, 503, 646 
\bibitem[Hopkins et al.(2001)]{2001AJ....122..288H} Hopkins, A.~M., 
Connolly, A.~J., Haarsma, D.~B., \& Cram, L.~E.\ 2001, \aj, 122, 288 
\bibitem[Hopkins 
\& Beacom(2006)]{2006ApJ...651..142H} Hopkins, A.~M., \& Beacom, J.~F.\ 2006, \apj, 651, 142 
\bibitem[Ilbert et 
al.(2005)]{2005A&A...439..863I} Ilbert, O., Tresse, L., Zucca, E., et al.\ 2005, \aap, 439, 863 
\bibitem[Ilbert et 
al.(2006)]{2006A&A...457..841I} Ilbert, O., Arnouts, S., McCracken, H.~J., et al.\ 2006, \aap, 457, 841
\bibitem[Ilbert et al.(2009)]{2009ApJ...690.1236I} Ilbert, O., et al.\ 
2009, \apj, 690, 1236 
\bibitem[Iwamuro et al. (2011)]{FIBRE-pac} Iwamuro, F., Moritani, 
Y., Yabe, K., et al.\ 2011, PASJ, 2011, in press
\bibitem[Kennicutt(1998)]{1998ARA&A..36..189K} Kennicutt, R.~C., Jr.\ 1998, \araa, 36, 189
\bibitem[Kennicutt et al.(2009)]{2009ApJ...703.1672K} Kennicutt, R.~C., 
Jr., et al.\ 2009, \apj, 703, 1672 
\bibitem[Kewley 
\& Dopita(2002)]{2002ApJS..142...35K} Kewley, L.~J., \& Dopita, M.~A.\ 2002, \apjs, 142, 35
\bibitem[Kewley 
\& Ellison(2008)]{2008ApJ...681.1183K} Kewley, L.~J., \& Ellison, S.~L.\ 2008, \apj, 681, 1183 
\bibitem[Kimura et al. (2010)], Kimura, M., et al., 2010, PASJ, 62, 5, 1135-1147.
\bibitem[Le Floc'h et al.(2009)]{2009ApJ...703..222L} Le Floc'h, E., 
Aussel, H., Ilbert, O., et al.\ 2009, \apj, 703, 222 
\bibitem[Levenson et al.(2010)]{2010MNRAS.409...83L} Levenson, L., et al.\ 
2010, \mnras, 409, 83 
\bibitem[Le F{\`e}vre et 
al.(2005)]{2005A&A...439..845L} Le F{\`e}vre, O., Vettolani, G., Garilli, B., et al.\ 2005, \aap, 439, 845 
\bibitem[Lilly et al.(1996)]{1996ApJ...460L...1L} Lilly, S.~J., Le Fevre, 
O., Hammer, F., \& Crampton, D.\ 1996, \apjl, 460, L1 
\bibitem[Lilly et al.(2007)]{2007ApJS..172...70L} Lilly, S.~J., et al.\ 
2007, \apjs, 172, 70 
\bibitem[Lonsdale et al.(2003)]{2003PASP..115..897L} Lonsdale, C.~J., et 
al.\ 2003, \pasp, 115, 897
\bibitem[Madau et al.(1996)]{1996MNRAS.283.1388M} Madau, P., Ferguson, 
H.~C., Dickinson, M.~E., Giavalisco, M., Steidel, C.~C., 
\& Fruchter, A.\ 1996, \mnras, 283, 1388
\bibitem[Maiolino et 
al.(2008)]{2008A&A...488..463M} Maiolino, R., Nagao, T., Grazian, A., et al.\ 2008, \aap, 488, 463 
\bibitem[Meurer et al.(1999)]{1999ApJ...521...64M} Meurer, G.~R., Heckman, 
T.~M., \& Calzetti, D.\ 1999, \apj, 521, 64 
\bibitem[McCracken et al.(2010)]{2010ApJ...708..202M} McCracken, H.~J., 
Capak, P., Salvato, M., et al.\ 2010, \apj, 708, 202 
\bibitem[McLure et al.(2010)]{2010MNRAS.403..960M} McLure, R.~J., Dunlop, 
J.~S., Cirasuolo, M., et al.\ 2010, \mnras, 403, 960 
\bibitem[Micha{\l}owski et al.(2011)]{2011arXiv1108.6058M} Micha{\l}owski, 
M.~J., Dunlop, J.~S., Cirasuolo, M., et al.\ 2011, arXiv:1108.6058 
\bibitem[Moustakas 
\& Kennicutt(2006)]{2006ApJS..164...81M} Moustakas, J., \& Kennicutt, R.~C., Jr.\ 2006, \apjs, 164, 81 
\bibitem[Oliver et al.(2012)]{HermesSurvey} Oliver, S., et al.\ 2012, \mnras, in press
\bibitem[Osterbrock 1989]{} Osterbrock, D.~E., 1989, Astrophysics of Gaseous Nebulae and Active Galactic Nuclei. University Science Books, Mill Valley, CA
\bibitem[Pettini 
\& Pagel(2004)]{2004MNRAS.348L..59P} Pettini, M., \& Pagel, B.~E.~J.\ 2004, \mnras, 348, L59 
\bibitem[Pilbratt et 
al.(2010)]{2010A&A...518L...1P} Pilbratt, G.~L., et al.\ 2010, \aap, 518, L1 
\bibitem[Reddy et al.(2010)]{2010ApJ...712.1070R} Reddy, N.~A., Erb, D.~K., 
Pettini, M., Steidel, C.~C., \& Shapley, A.~E.\ 2010, \apj, 712, 1070
\bibitem[Rieke et al.(2009)]{2009ApJ...692..556R} Rieke, G.~H., 
Alonso-Herrero, A., Weiner, B.~J., P{\'e}rez-Gonz{\'a}lez, P.~G., Blaylock, 
M., Donley, J.~L., \& Marcillac, D.\ 2009, \apj, 692, 556 
\bibitem[Roseboom et al.(2010)]{2010MNRAS.409...48R} Roseboom, I.~G., et 
al.\ 2010, \mnras, 409, 48 
\bibitem[Roseboom et al. (2012)]{2012MNRAS.419.2758R} Roseboom, I.~G., et al.\ 2012, \mnras, 419, 2758
\bibitem[Salpeter(1955)]{1955ApJ...121..161S} Salpeter, E.~E.\ 1955, \apj, 
121, 161 
\bibitem[Sanders et al.(2007)]{2007ApJS..172...86S} Sanders, D.~B., et al.\ 
2007, \apjs, 172, 86 
\bibitem[Sarzi et al.(2006)]{2006MNRAS.366.1151S} Sarzi, M., et al.\ 2006, 
\mnras, 366, 1151 
\bibitem[Schiminovich et al.(2005)]{2005ApJ...619L..47S} Schiminovich, D., 
Ilbert, O., Arnouts, S., et al.\ 2005, \apjl, 619, L47 
\bibitem[Scoville et al.(2007)]{2007ApJS..172....1S} Scoville, N., Aussel, 
H., Brusa, M., et al.\ 2007, \apjs, 172, 1 
\bibitem[Seymour et al.(2008)]{2008MNRAS.386.1695S} Seymour, N., Dwelly, 
T., Moss, D., et al.\ 2008, \mnras, 386, 1695 
\bibitem[Skrutskie et al.(2006)]{2006AJ....131.1163S} Skrutskie, M.~F., 
Cutri, R.~M., Stiening, R., et al.\ 2006, \aj, 131, 1163 
\bibitem[Swinyard et al.(2010)] {SPIREcalibration} Swinyard, B., et al.\ 2010, \aap, 518, 4
\bibitem[Totani et al. (2011)]{2011totani} Totani, T., Takeuchi, T.~T., Nagashima, M., Kobayashi, M.~A.~.R., Makiya, R., 2011, PASJ, in press (arXiv:1103.5402)
\bibitem[Tremonti et al.(2004)]{2004ApJ...613..898T} Tremonti, C.~A., 
Heckman, T.~M., Kauffmann, G., et al.\ 2004, \apj, 613, 898 
\bibitem[Wijesinghe et al.(2011)]{2011MNRAS.415.1002W} Wijesinghe, D.~B., 
et al.\ 2011, \mnras, 415, 1002
\bibitem[Yabe et al.(2012)]{2012PASJ} Yabe, K., et al., 2012, PASJ, 64, 60.
\bibitem[Zahid et al.(2011)]{2011ApJ...730..137Z} Zahid, H.~J., Kewley, 
L.~J., \& Bresolin, F.\ 2011, \apj, 730, 137 
\end{thebibliography}
\end{document}